\documentclass{article}
\usepackage{graphicx}
\begin{document}

\begin{center}
{\bf \LARGE Spectroscopic evidence of Multiple Stellar Populations in Globular 
Clusters}\footnote{Invited review talk given on August, 10 2016 at the Conference 
"Star Clusters: From Infancy to Teenagehood", Heidelberg, organized by
Genevi\`eve Parmentier.}
\bigskip

{\Large Eugenio Carretta \\
\bigskip
INAF-Osservatorio Astronomico di Bologna; email eugenio.carretta@oabo.inaf.it} \\

\bigskip
ABSTRACT \\
\end{center}

Galactic globular clusters are not simple stellar populations. And nothing
is simple in their study, basically because we try to reconstruct chains
of events that occurred at redshift $z > 2-3$ by observing these objects
at $z=0$, after a Hubble time.
Fortunately,  spectroscopy offers a magnifying lens: differences of
tens or hundreds of Myrs between stellar generations are translated into
differences in abundances up to a full dex.
I will review the complex pattern emerging by the combined efforts of
different groups, focusing on the chemical signatures of multiple populations
in globular clusters.\\


\begin{figure}
\begin{center}
\includegraphics[scale=0.4]{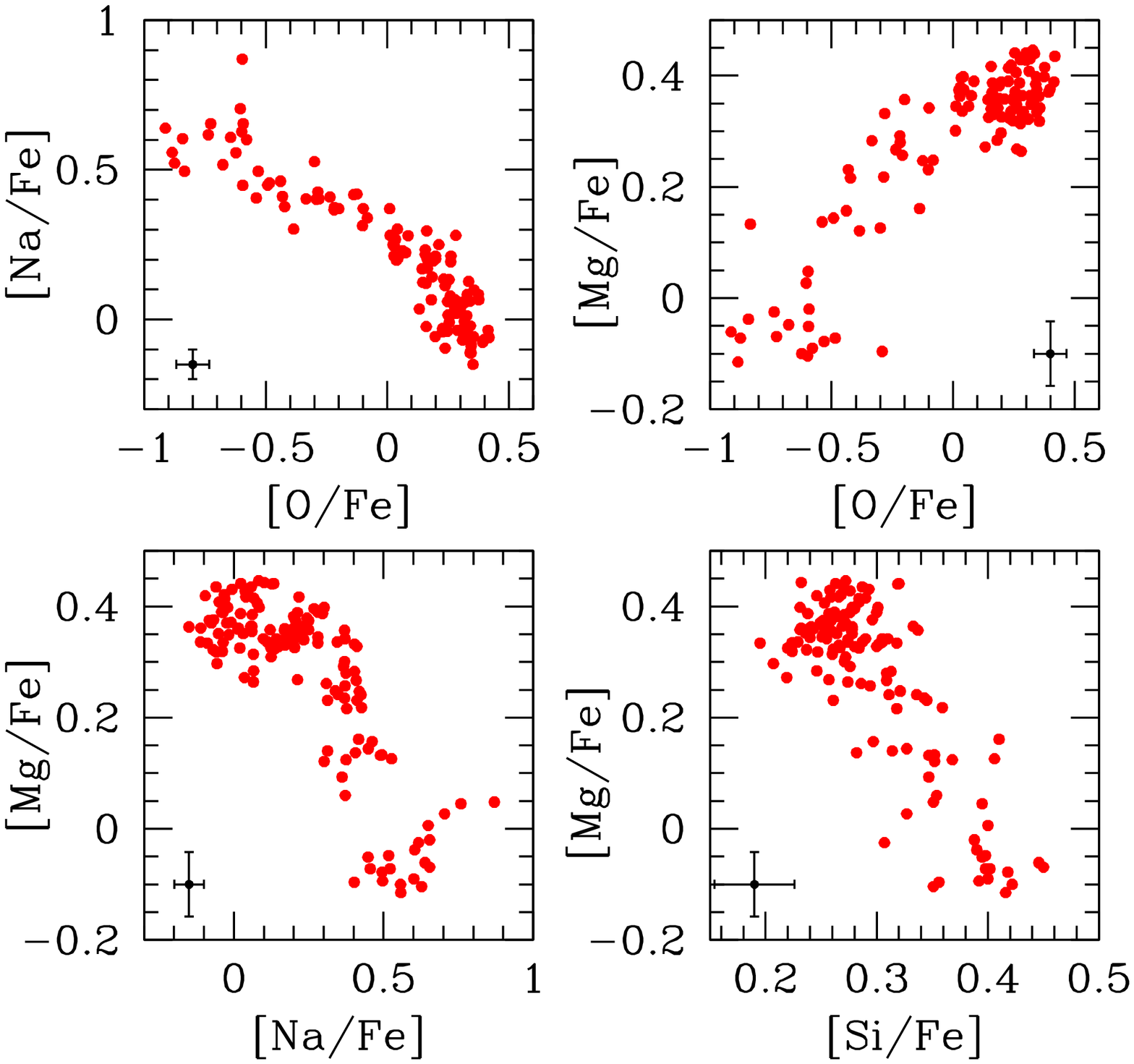}
\includegraphics[scale=0.4]{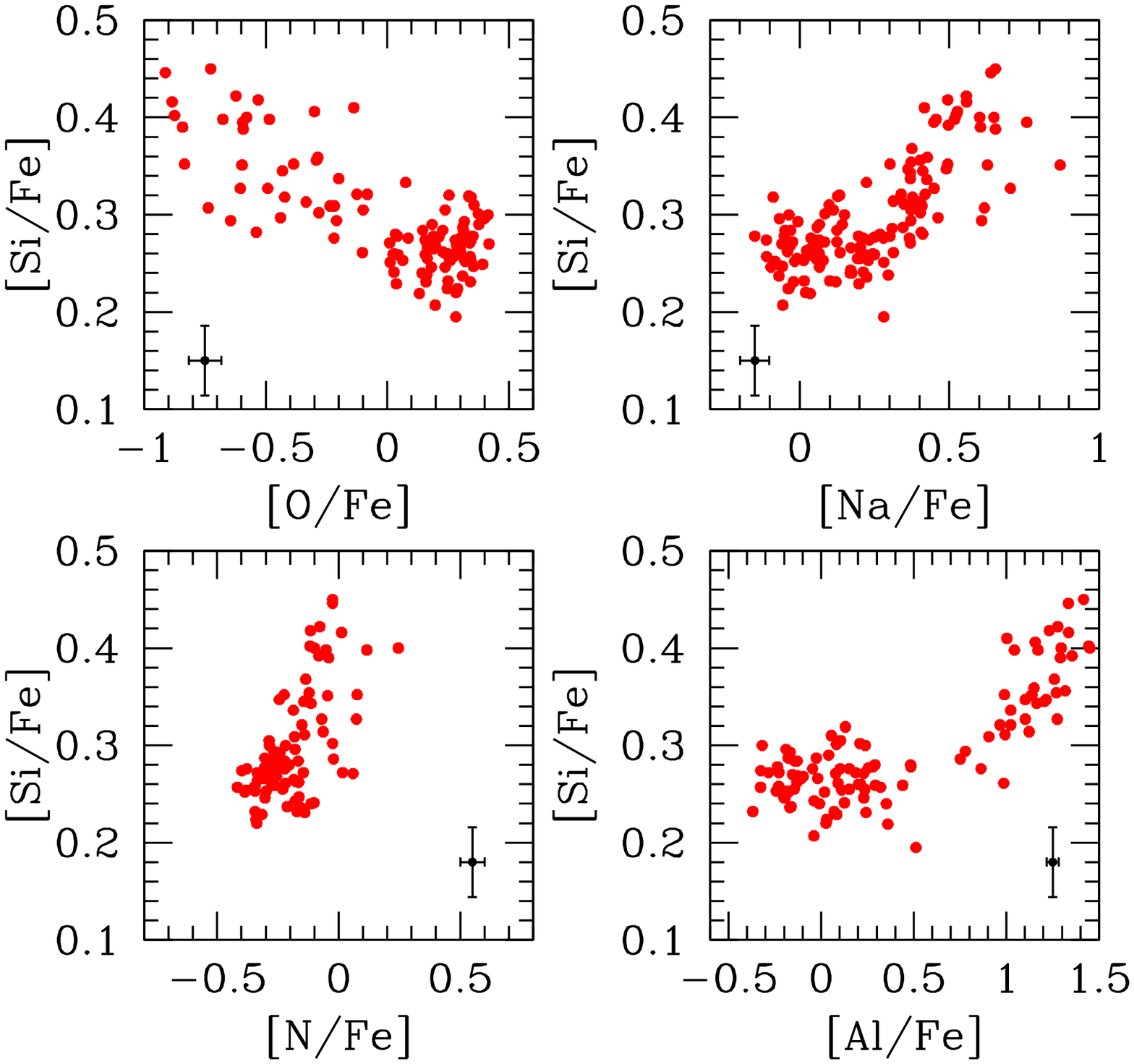}
\includegraphics[scale=0.4]{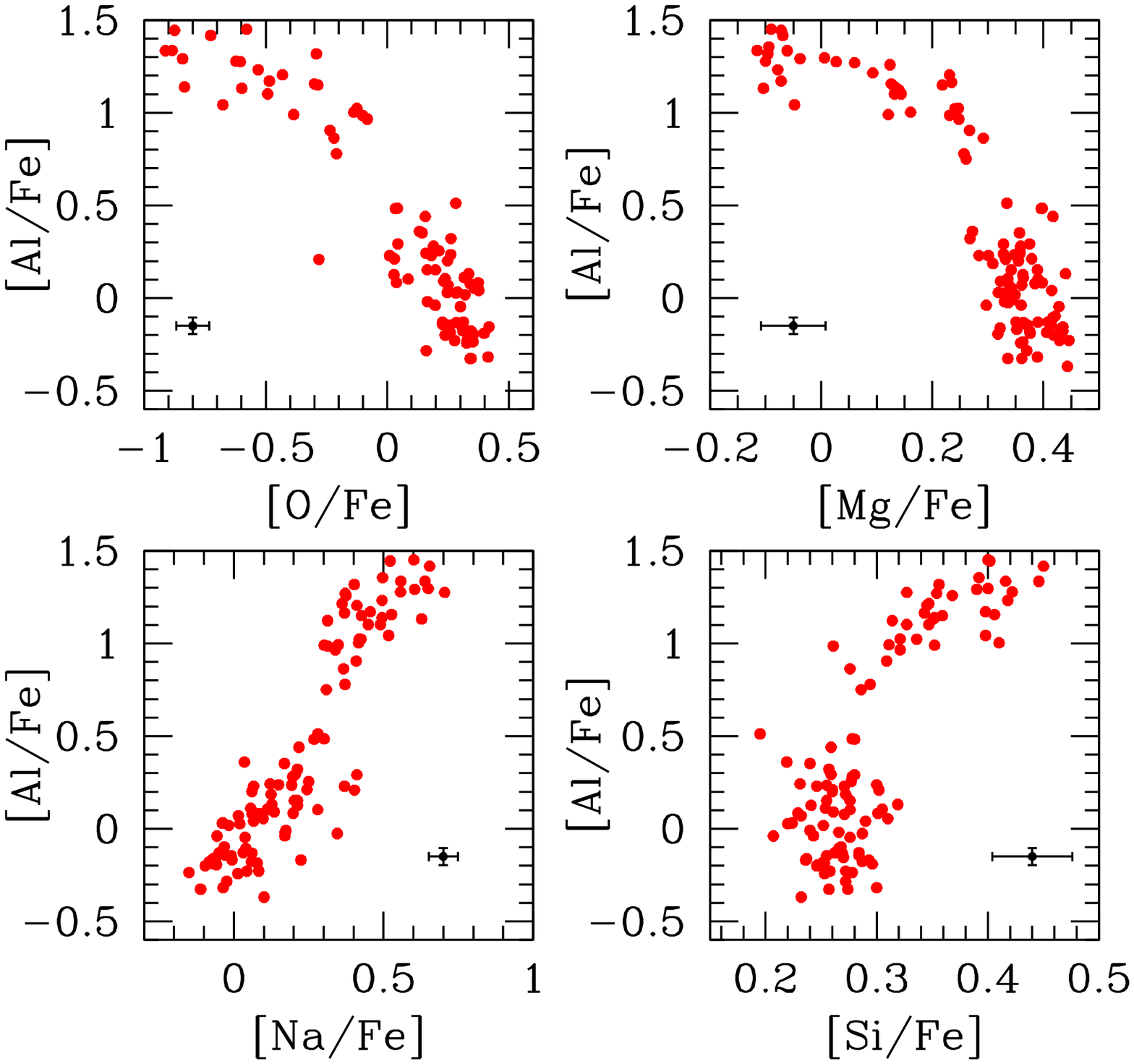}
\includegraphics[scale=0.4]{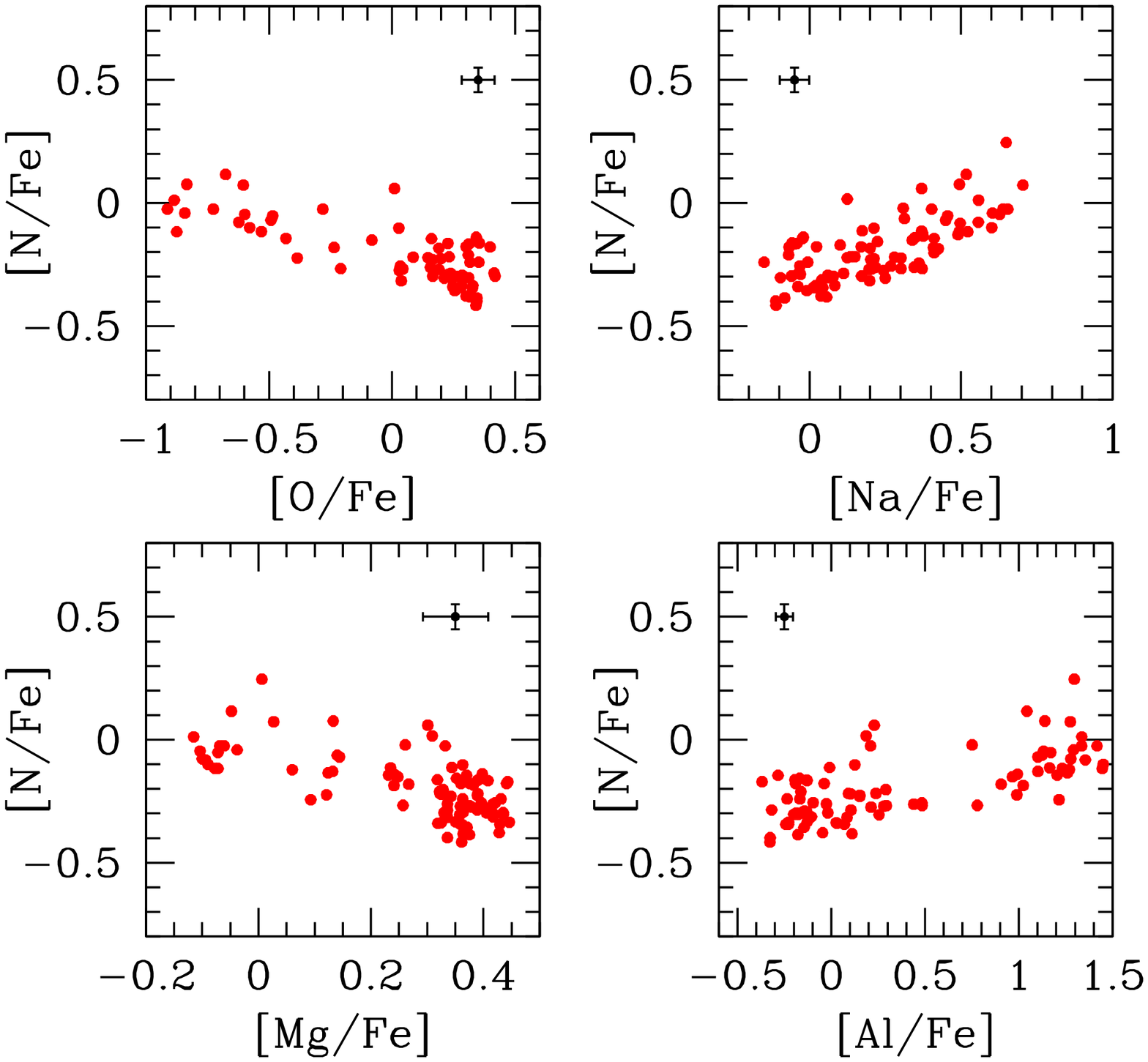}
\caption{Trends of abundance ratios for light elements in RGB stars of
NGC~2808. O, Na, Si, Mg are from Carretta (2015), Al and N abundances from
Carretta et al. (in prep.).}
\end{center}
\end{figure}

\section{Introduction}
This is a mixed audience, with different expertises, so I will give a general
overview, focussing on observations, but why spectroscopy? There are
two main reasons: one, it is the first and most direct evidence, following from 
the very same definition of $simple$ stellar populations: coeval stars with the
same initial chemical composition. Therefore,  if we see abundance variations at
the same evolutionary phase, then we see multiple stellar populations.
Photometry is, literally, a $filtered$ indicator, showing multiple populations
when selected bandpasses,
sensitive to CNO elements, are used. However, no heavy elements (Na, Mg, Al) 
are detected using wide and intermediate band filters.
Of course, there are also benefits (large samples, radial distributions, helium
estimates, etc, see e.g. Milone et al. 2012, Larsen et al. 2014), but they will be
discussed elsewhere.

The second reason is that we observe now in globular clusters (GCs) sub-solar 
stellar masses, whose main sequence lifetime is comparable to a Hubble time.
So globular clusters are very old, born at high redshift and it is challenging
to understand how they formed by looking at the $z=0$ end-products.
But fortunately spectroscopy works like a magnifying lens.
After the first burst, second generation stars may form after only a few million
or a few ten of million years, but even these tiny age differences 
translate into huge variations in the content
of a few elements, up to 1 full dex, providing an excellent time resolution:
we can easily distinguish primordial stars with only supernovae
nucleosynthesis from younger generations with clear traces of self-enrichment.

In this talk I will present an overview on what we see and where, in a globular
cluster  and among clusters; what is the mechanism and what are the
links with global cluster properties.

\section{What we see}
By comparing the ranges of element abundances in field and cluster stars, it is well known that 
very large variations (at the same metallicity) are restricted to the  dense environment of globular clusters
and only concern light elements. How do these elements vary?

As an example one can use NGC~2808, one of the best studied clusters, with many
studies on the involved elements in several evolutionary phases: Carretta et al.
(2003: RGB), Moehler et al. (2004: blue hook), Carretta et al. (2004: RGB),
Carretta (2006: RGB), Pace et al. (2006: BHB),
Bragaglia et al. (2010a: TO), Pasquini et al. (2011: RGB), Gratton et al. (2011:
HB), Marino et al. (2014: HB), Carretta (2014: RGB), Mucciarelli et al. (2015:
RGB), D'Orazi et al. (2015: RGB), Carretta (2015: RGB).
By using the same sample for all plots, the typical relations among abundance
ratios of light elements in NGC~2808 are shown in Figures 1 and 2. These trends
involve Na, O (the famous anticorrelation discovered by the Lick-Texas group,
see Kraft 1994), Mg, N, Si, Al and, finally, the heaviest elements included:
K and Sc.

\begin{figure}
\begin{center}
\includegraphics[scale=0.4]{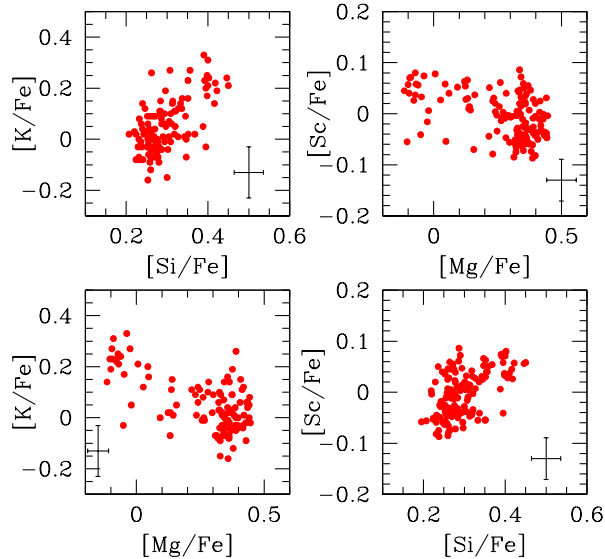}
\caption{as in previous figure for Mg, Si, Sc from Carretta (2015) and 
K abundances from Mucciarelli et al. (2015).}
\end{center}
\end{figure}

All these elements are not in scatter plots, but some of them decrease and 
simultaneously some increase from the level established by supernovae, in second
generation stars, so that  the chemical signature we see are well defined
(anti)correlations among these light elements.

\bigskip
Do we know a common process able to explain all the observed patterns? Yes: the
origin of these signature was identified in proton capture reactions in
hydrogen burning at high temperature, where several synthesis chains are
simultaneously active (Denisenkov \& Denisenkova 1989, Langer et al. 1993).
However, the required temperatures are high ($> 40 \times 10^6$ K for the ON and
NeNa cycles, and even higher for the MgAl cycle, $T> 70 \times 10^6$ K), not reached
in the interior of the presently observed GC low mass stars. 
The implication is that this nuclear burning occurred in stars more massive than
those evolving today, and this is the reason why
since Gratton et al. (2001) discovered the Na-O and Mg-Al anticorrelations among
unevolved GC stars (Figure 3), these (anti)correlations {\it are simply an alias for
multiple stellar populations.}

\begin{figure}
\begin{center}
\includegraphics[scale=0.35]{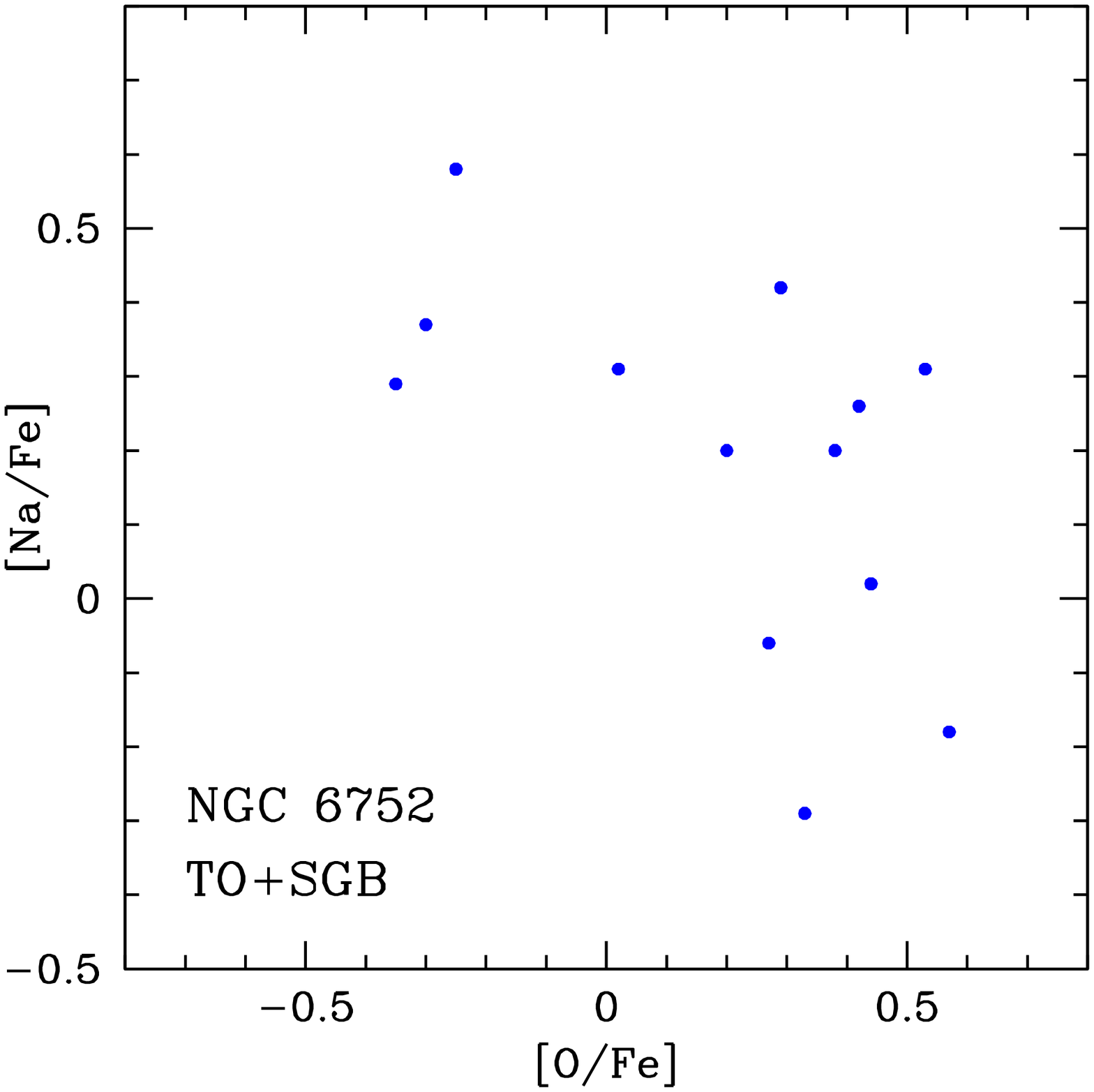}
\includegraphics[bb=18 149 554 615, clip, scale=0.45]{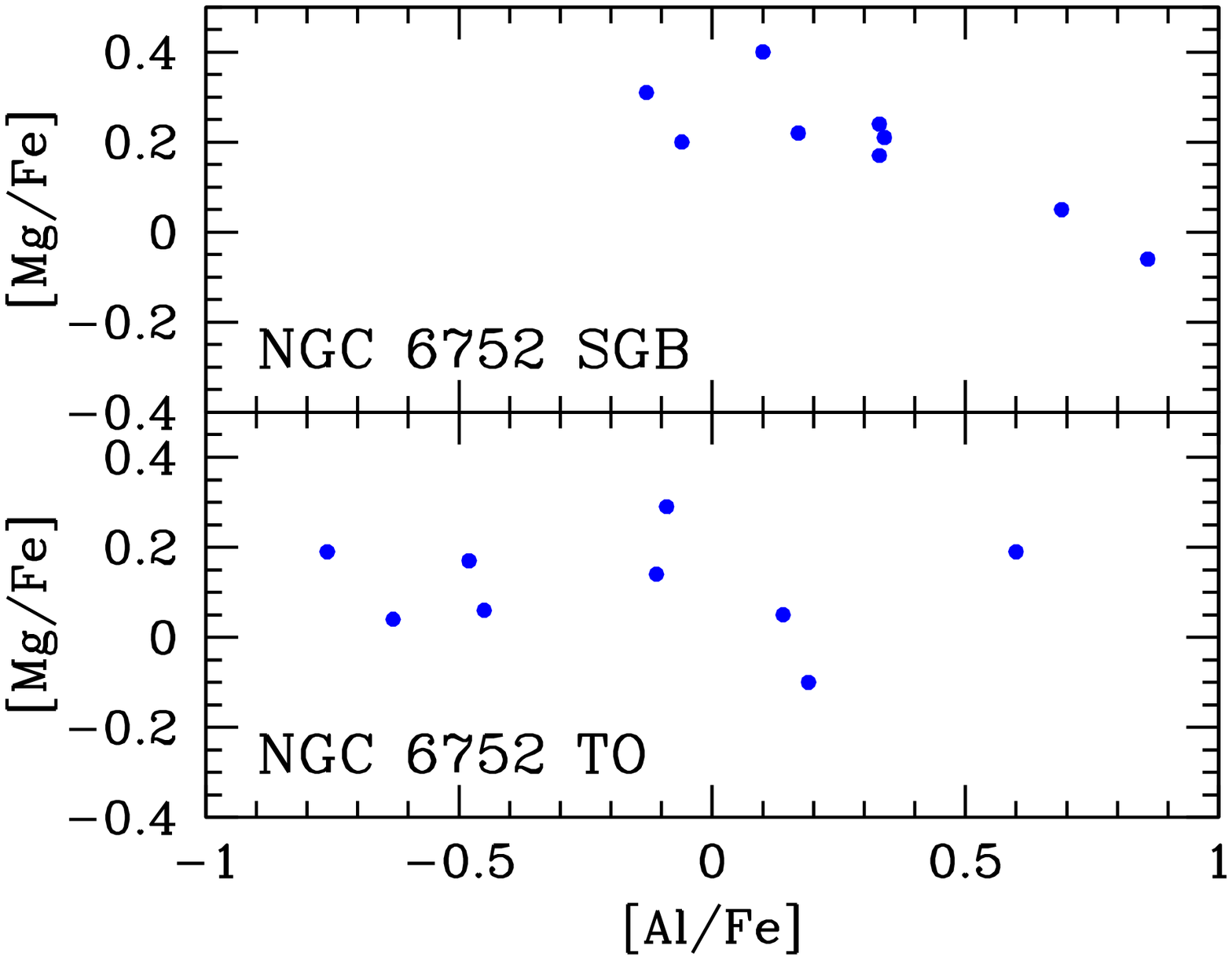}
\caption{Na-O (left) and Mg-Al (right) anticorrelations in unevolved stars 
of NGC~6752 from Gratton et al. (2001).}
\end{center}
\end{figure}

The temperature stratification may provide in principle useful constraints  on
the masses of the involved producers: Prantzos et al. (2007) were able to
pinpoint a temperature range where all the observations in NGC~6752 were
simultaneously satisfied by proton-capture reactions. Several classes of
possible polluters fit in this range (massive AGB and super-AGB stars, Ventura et
al. 2001; fast rotating massive stars, Decressin et al. 2007; massive binaries,
de Mink et al. 2009; supermassive stars, Denissenkov and Hartwick 2014), but the
common scenario for feedback-regulated secondary star formation is that second
generation (SG) stars are formed by nuclear ejecta processed in the most massive
first generation (FG) stars, diluted with different amounts of unprocessed gas,
generating the observed anticorrelations.

Dilution with gas of primordial composition is badly required for several
reasons (i) the star formation is not 100\% efficient (e.g. Lada \&
Lada 2003), so a fraction of gas is left over; (ii) 
the chemical feedback from a star is no more than a few 
percent of its mass, not enough to form all the SG assuming a normal IMF (e.g.
de Mink et al. 2009); (iii) dilution is mandatory to obtain a Na-O
anticorrelation from AGB stars, where Na and O are correlated (e.g. D'Ercole et
al. 2008, 2010, 2011); and finally (iv)
we see in second generation stars traces of the fragile litium which is
destroyed at lower temperatures than those involved in hot H-burning
(although it can be also produced in AGB stars, see D'Antona et al. 2012).

\subsection{A genetic link between stellar generations?}
There seems to be a recent fashion to deny a direct genetic link between first and second
generation stars (e.g., the group by Bastian and collaborators, or Schiavon et al.
2016). In other words, what we see would not be nucleosynthesis, due to problems still
existing in the theoretical scenarios (yields, interplay between helium and the
proton-capture elements, population ratios, to quote a few).
To clarify this issue, let us take the observer's route.

\begin{itemize}
\item We know that in GC stars carbon and nitrogen abundances are anticorrelated, but
as C decreases the sum C+N increases (see Figure 4, left panel). Thus, either O
is also involved (being transformed into N), or we are seeing variable amounts
of nitrogen. However, the sum C+N+O is found to be pretty constant in most
globular cluster stars (Figure 4, right panel; see also results from APOGEE,
Meszaros et al. 2015). Hence, the complete CNO cycle is accounted for.

\begin{figure}
\begin{center}
\includegraphics[scale=0.40]{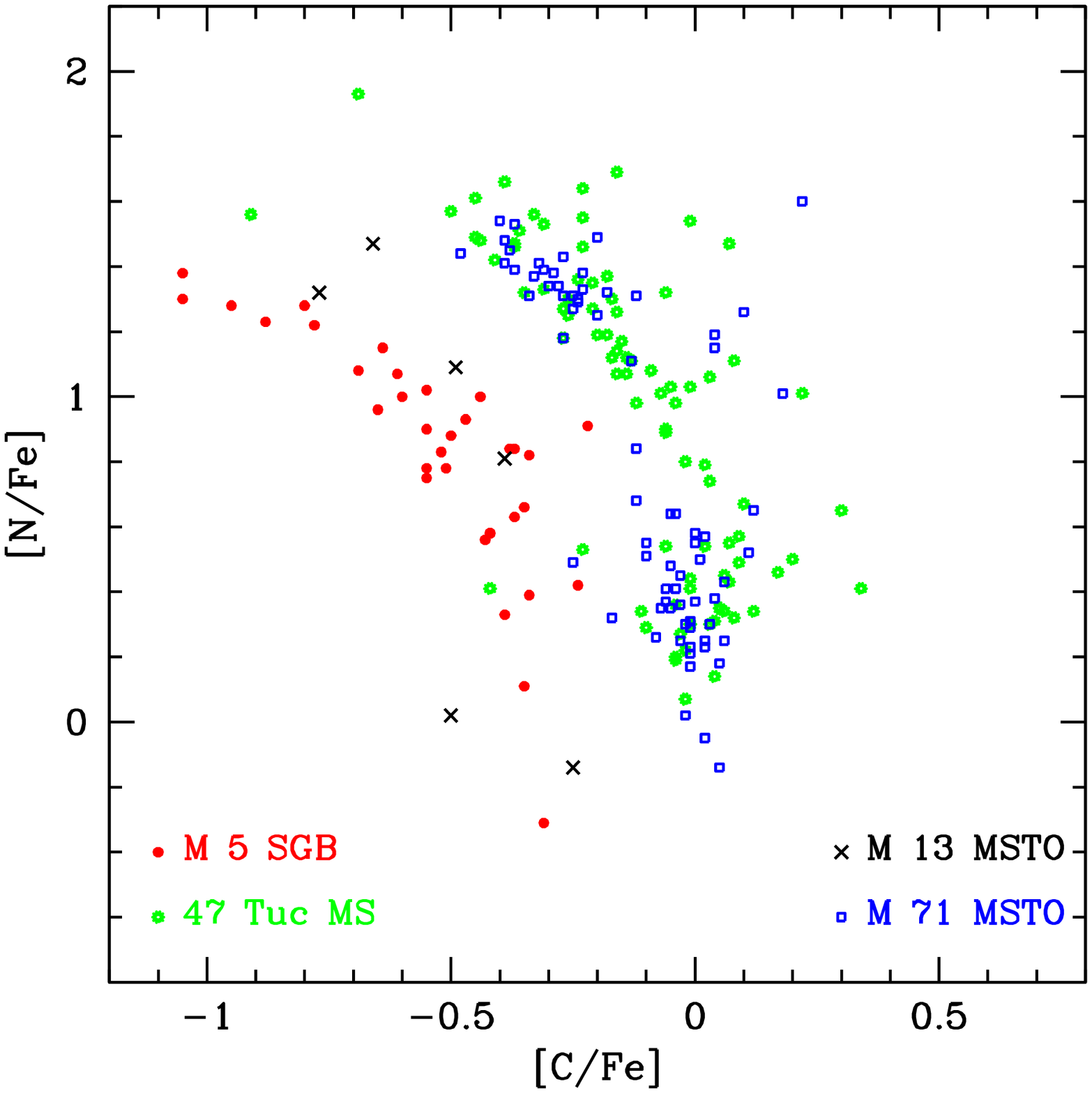}
\includegraphics[scale=0.40]{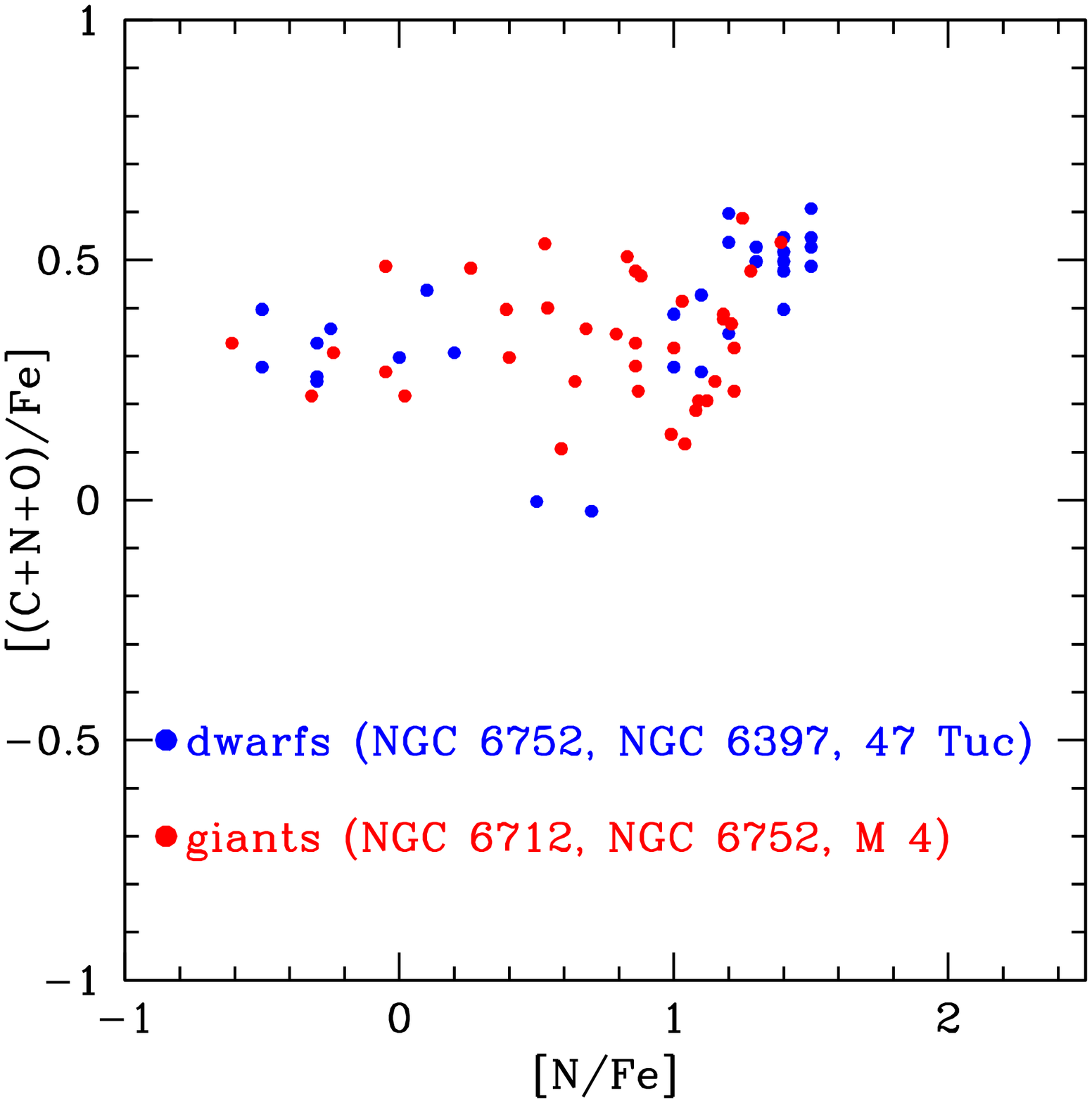}
\caption{On the left, the [C/Fe] ratios as a function of the sum [(C+N)/Fe] for unevolved
stars in M~5, M~13, M~71 and 47~Tuc from Briley \& Cohen (2001), Cohen, Briley,
Stetson (2002) and Briley et al. (2004). On the right: C+N+O sum as a function
of [N/Fe] ratios for dwarfs in NGC~6752, NGC~6397, and 47 Tuc (Carretta et al.
2005) and for giants in NGC~6712 (Yong et al. 2008), NGC~6752 (Yong et al. 2015),
and M~4 (Ivans et al. 1999, Smith et al. 2005).}
\end{center}
\end{figure}

\item The stars with enhanced sodium and depleted oxygen are also rich in 
aluminum (see Figure 1): the Na-Al correlation links the NeNa and MgAl cycles
and moreover the sum Mg+Al does not vary (e.g. Meszaros et al. 2015). Hence,
the NeNa and MgAl cycles are accounted for.

\item In metal-poor and massive clusters we see silicon anticorrelated with 
magnesium and/or alternatively correlated with aluminum: results by Carretta et
al. (2009) are shown in Figure 5, confirming those for NGC~6752 by Yong et al.
(2005), and confirmed by the findings of Meszaros et al. (2015). Hence, the leakage of Mg 
on silicon (Karakas et al. 2003), bypassing production of Al, is accounted for 
in these clusters.

\begin{figure}
\begin{center}
\includegraphics[scale=0.50]{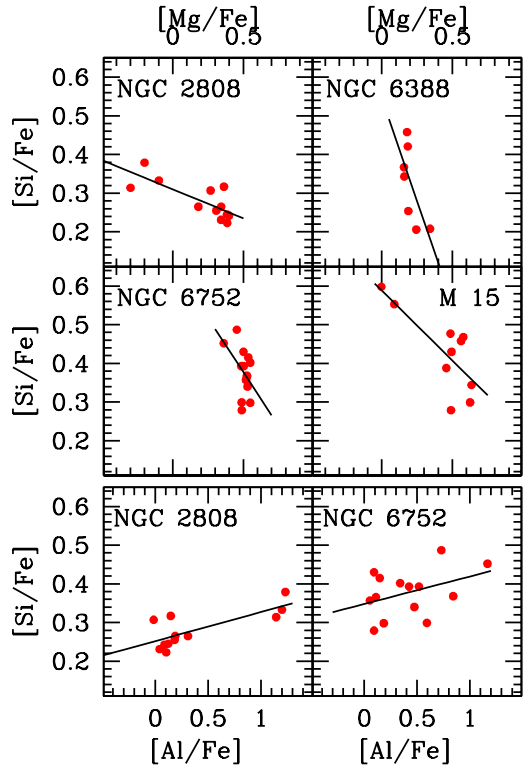}
\caption{Si-Mg anticorrelations and Si-Al correlations for massive and/or
metal-poor GCs from Carretta et al. (2009b).}
\end{center}
\end{figure}

\item In particular cases also star to star variations in heavier elements like K, Sc
are observed (see Figure 2), and this is explained when proton-captures on
argon, at very high temperature, produces these elements, bypassing Al
production (e.g. Ventura et al. 2012).
\end{itemize}

In summary, whatever is the proton-capture reaction involved, either destroying
or producing a given specise, we may observe the ashes of the nuclear burning in
second generation stars. 
Therefore, the Occam's razor tells us that all the observed pattern must be
nucleosynthesis: until a better alternative is proposed and motivated,
simultaneously explaining $all$ the relations among light-elements, we conclude
that we are seeing the signature of hot H-burning in stars more massive than
those observed today.

\section{Where do we see the spectroscopic evidence in GCs?}
By using the main chemical signature, the Na-O anticorrelation, we can ask in
what evolutionary phase multiple populations are observed in GCs. In a word, 
everywhere :
\begin {itemize}
\item
on  the main sequence (MS, see Figure 6, left panel), where the anticorrelation
is similar in extent to that observed along the RGB (apart from offsets due to
the analyses), despite very different stellar structure (negligible convective
envelopes in MS) and H-burning process (p-p in core burning on the MS and
CNO-shell burning along the RGB).

\begin{figure}
\begin{center}
\includegraphics[bb=19 148 570 438, clip, scale=0.70]{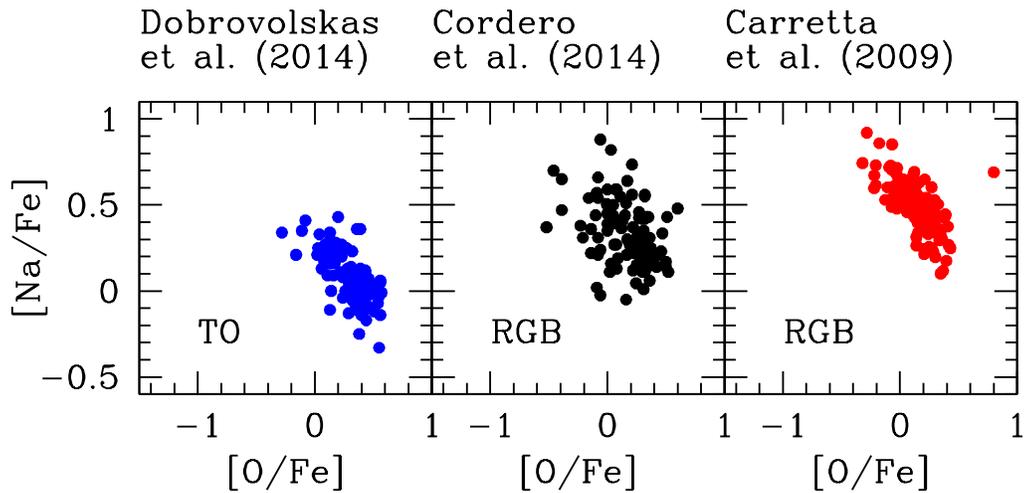}
\caption{Na-O anticorrelation in 47 Tuc for main sequence/turnoff stars from
Dobrovolskas et al. (left panel) and for RGB stars from Cordero et al. (2014) and
Carretta et al. (2009), central and right panels, respectively.}
\end{center}
\end{figure}

\item of course on the red giant branch, where large statistics allows to quantify
efficiently multiple populations as in our extensive and homogeneous FLAMES 
survey (Figure 7: Bragaglia et al. 2015; Carretta 2013, 2014, 2015; Carretta et
al. 2006, 2007a,b,c, 2009a,b,c, 2010a,b,c,d,e, 2011, 2012a,b,c, 2013a,b,
2014a,b,c, 2015; Gratton et al. 2006, 2007) and in more eterogeneous studies
(e.g. Johnson and Pilachowski 2012, Lind et al. 2009, Marino et al. 2008, Yong
et al. 2005, see Figure 8), with some useful northern addition from APOGEE
(Meszaros et al. 2015).

\begin{figure}
\begin{center}
\includegraphics[scale=0.6]{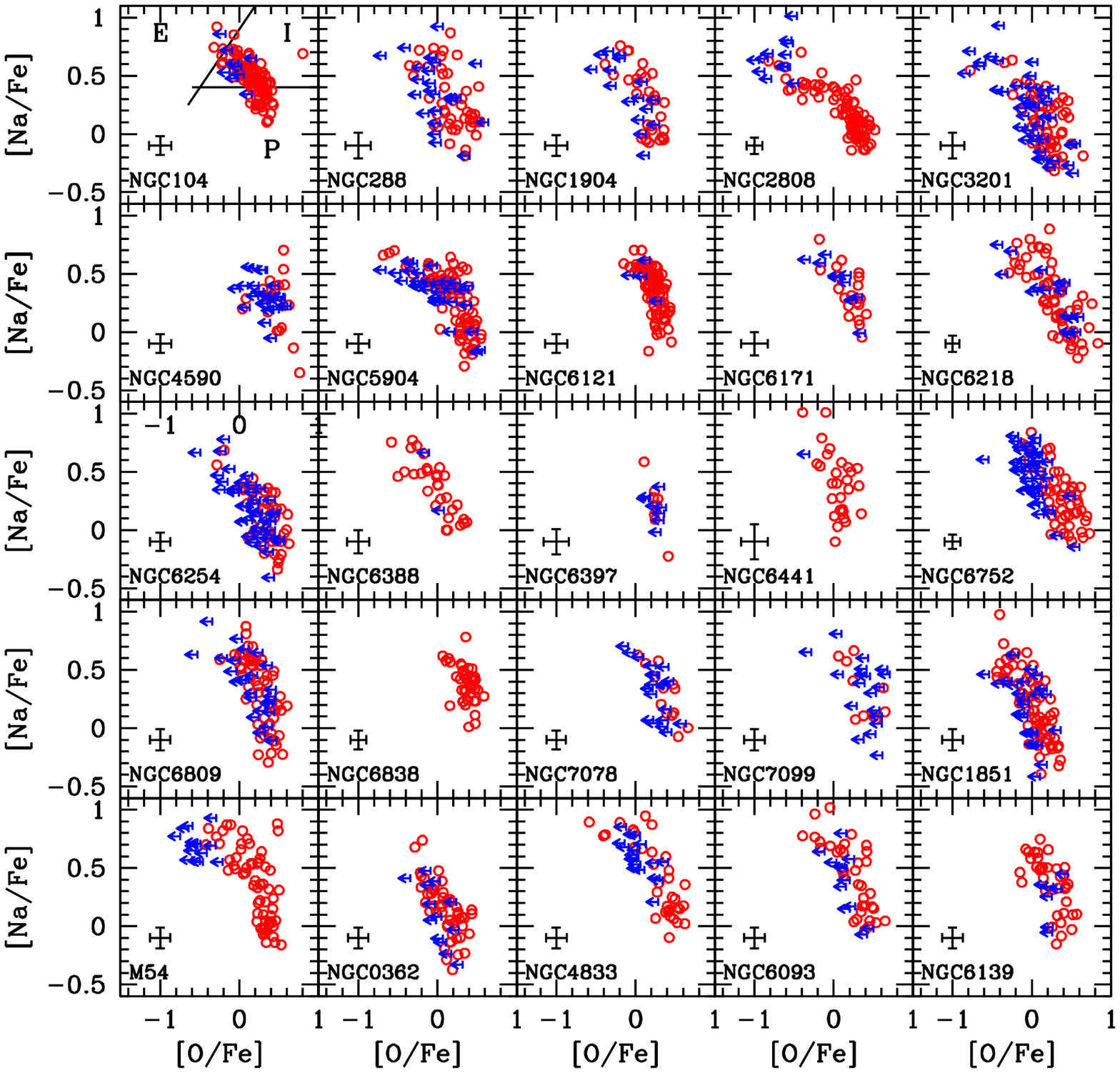}
\caption{Na-O anticorrelation in 25 GCs from our homogeneous FLAMES survey (see
text for references).}
\end{center}
\end{figure}

\begin{figure}
\begin{center}
\includegraphics[scale=0.3]{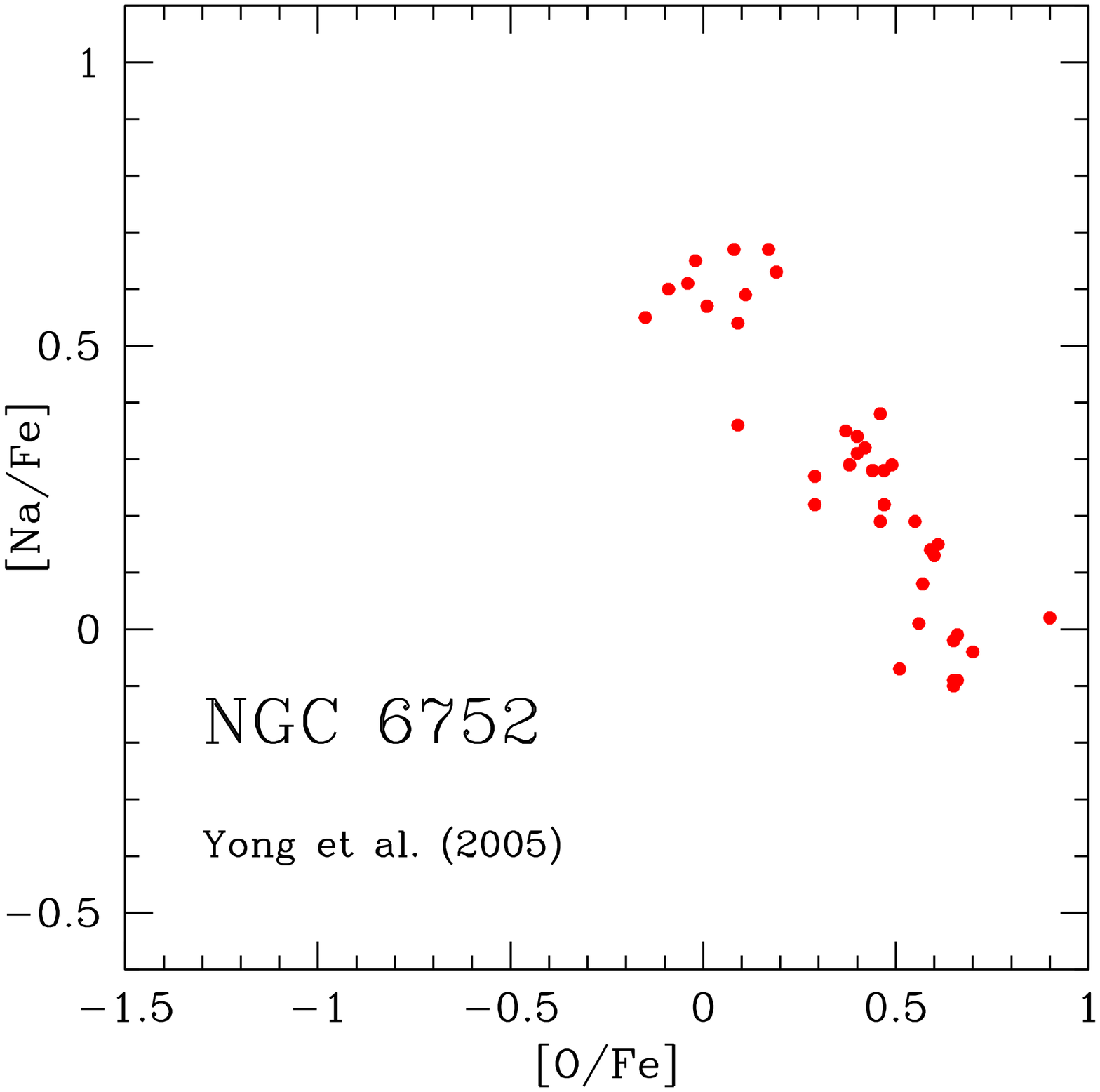}
\includegraphics[scale=0.3]{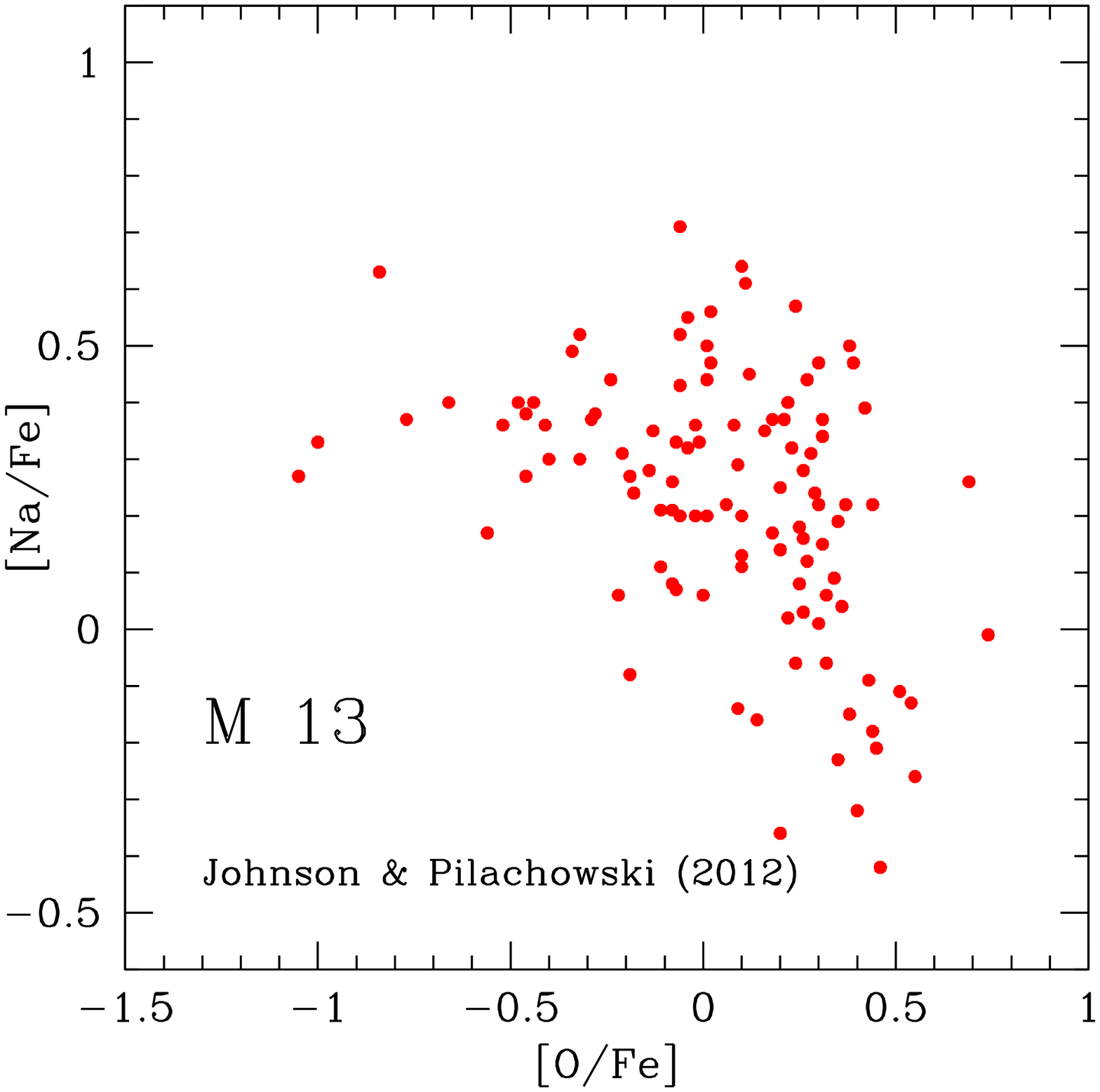}
\caption{Na-O anticorrelation in RGB stars of NGC~6752 (Yong et al. 2005, left
panel) and M~13 (Johnson and Pilachowski 2012, right panel).}
\end{center}
\end{figure}

\item on the horizontal branch (HB), where the extension of the Na-O anticorrelation
is similar to that measured on the RGB (Figure 9). However, Na-poor stars
are segregated on the red HB and Na-rich stars on the blue HB (Figure 10,
see also Gratton et al. 2011, 2013, 2014, 2015, Marino et al. 2011, Villanova et
al. 2009, 2012) because of different helium from nuclear processing, resulting
in slightly lower masses on the RGB (see e.g. D'Antona et al. 2002).
In fact, we know from both spectroscopy and photometry that second generation 
Na-rich stars are also enhanced in helium from  direct measurements 
(e.g. Pasquini et al. 2011, Villanova et al. 2012) or from a brighter bump on the
luminosity function of the RGB (Bragaglia et al. 2010b), as predicted by Salaris
et al (2006) for He-enhanced populations.

\begin{figure}
\begin{center}
\includegraphics[bb=19 146 586 512,clip, scale=0.40]{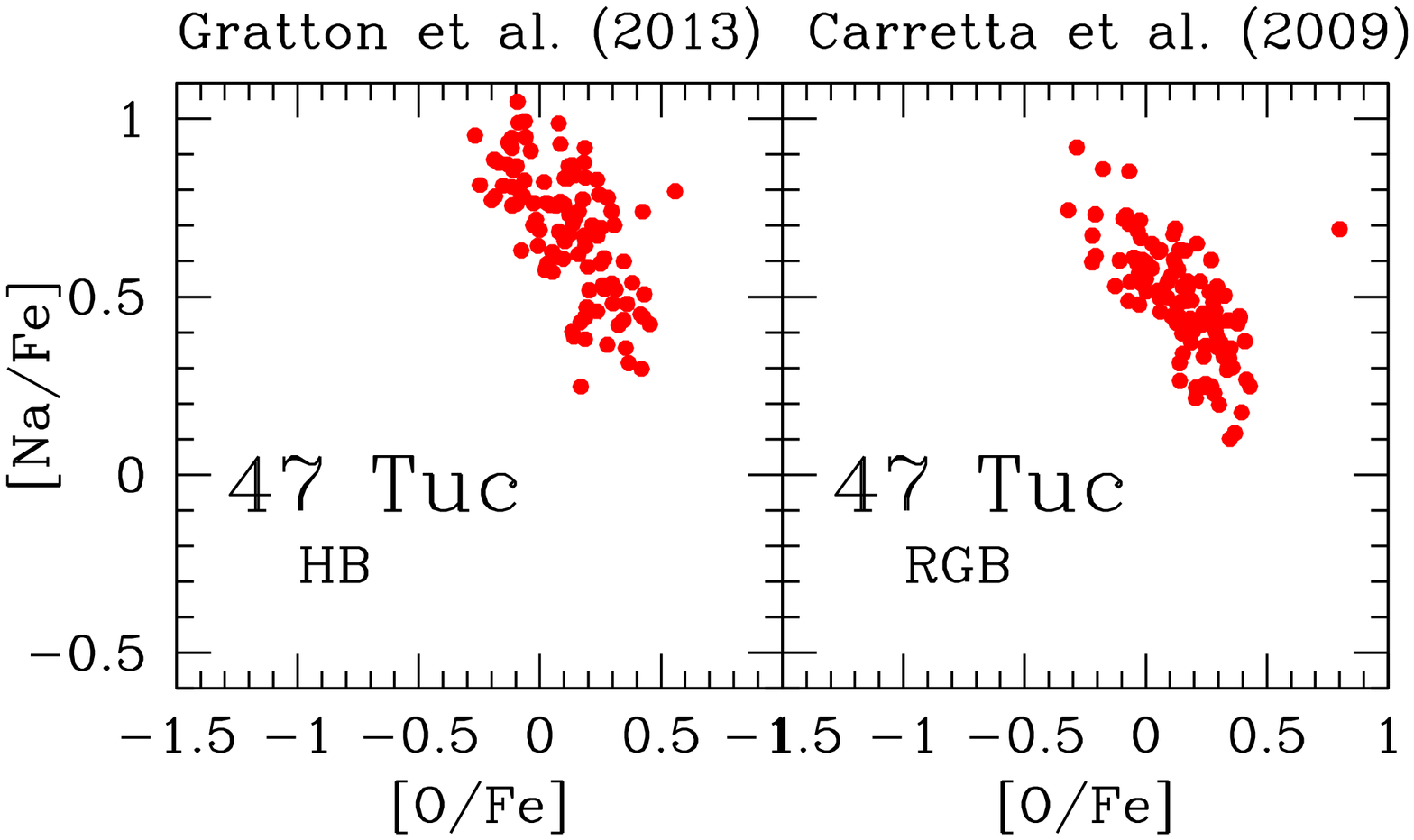}
\includegraphics[bb=19 146 586 512,clip, scale=0.40]{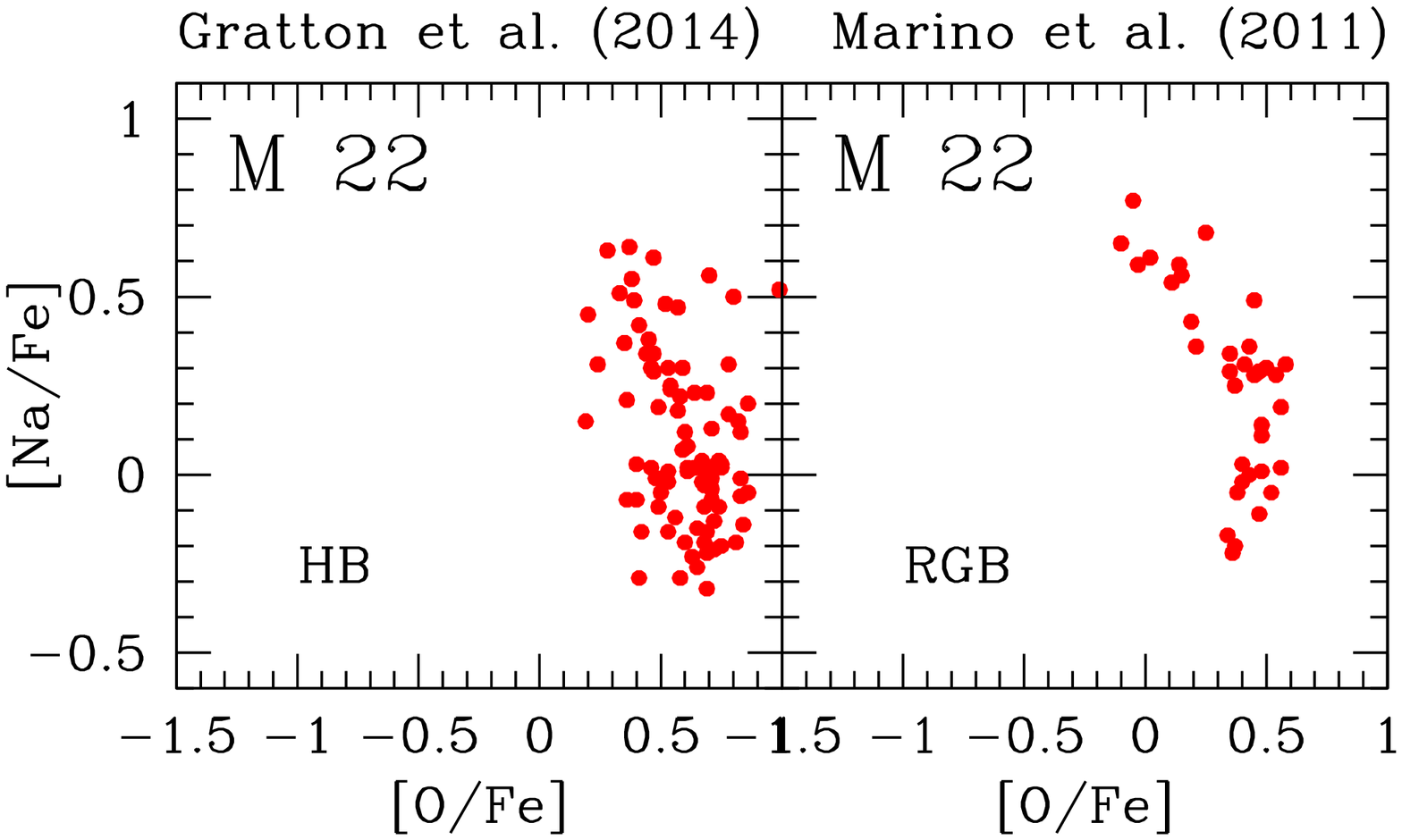}
\caption{Na-O anticorrelation in HB and RGB stars of 47 Tuc and M~22. References
are given on top.}
\end{center}
\end{figure}

\item the Na-O anticorrelation is also observed on the asymptotic giant branch (AGB),
even if some studies seem to differ (see Fig. 10).
Some recent works (Campbell et al. 2013, Johnson et al. 2015a, Lapenna et al. 2015,
2016, MacLean et al. 2016, Wang et al. 2016) find few or no second
generation stars on the AGB in some clusters (but see
Garcia-Hernandez et al. 2015 for a different view). 
\end{itemize}

\begin{figure}
\begin{center}
\includegraphics[scale=0.40]{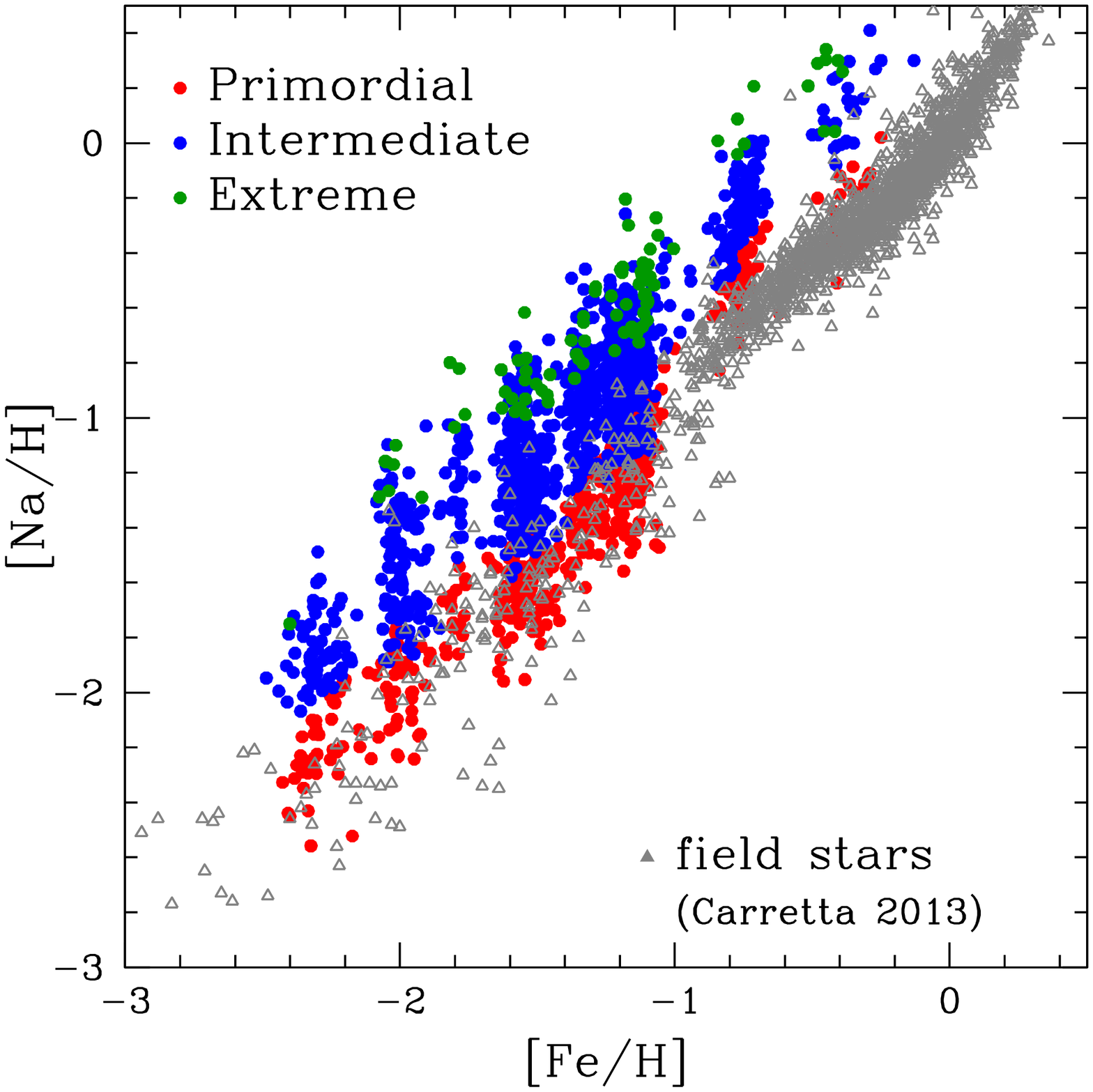}
\includegraphics[scale=0.40]{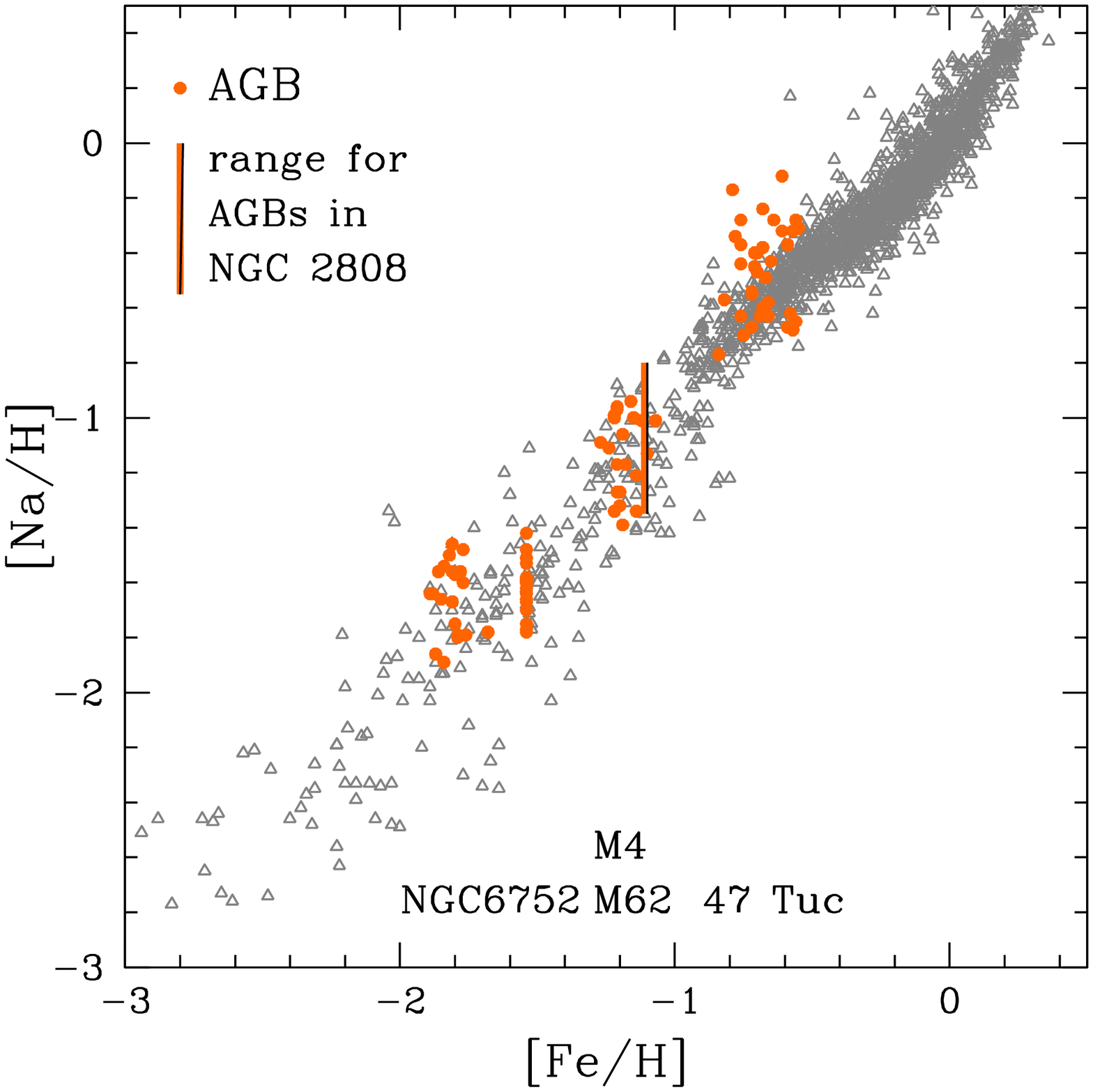}
\caption{[Na/H] ratio as a function of metallicity for Galactic field stars in
the solar neighborhood from Carretta (2013; empty grey triangles), with
superimposed RGB stars in 25 GCs of our FLAMES survey (left panel). In the right
panel, the same with superimposed AGB stars from several studies (see text for
references).}
\end{center}
\end{figure}

\begin{figure}
\begin{center}
\includegraphics[scale=0.40]{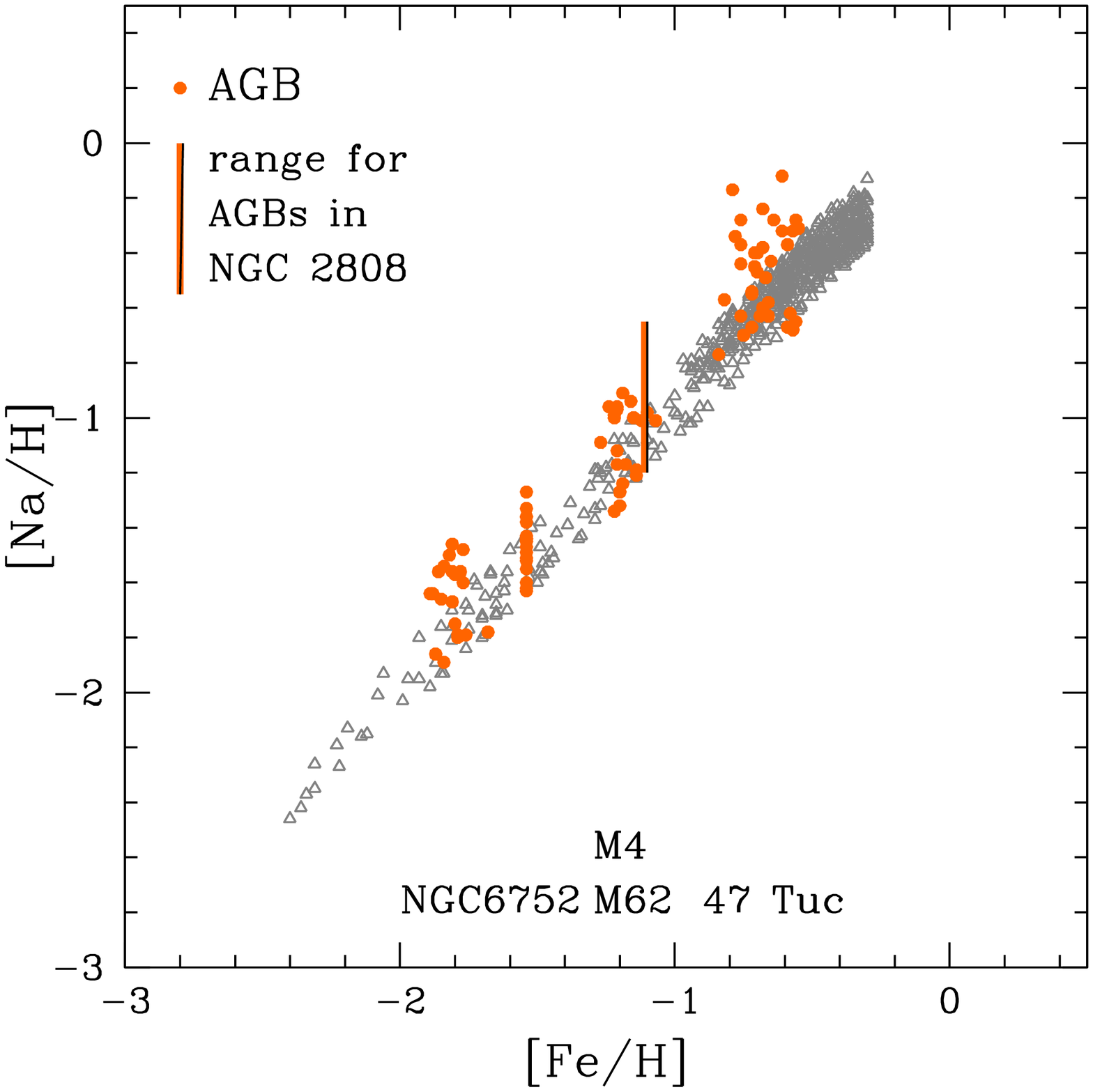}
\includegraphics[scale=0.40]{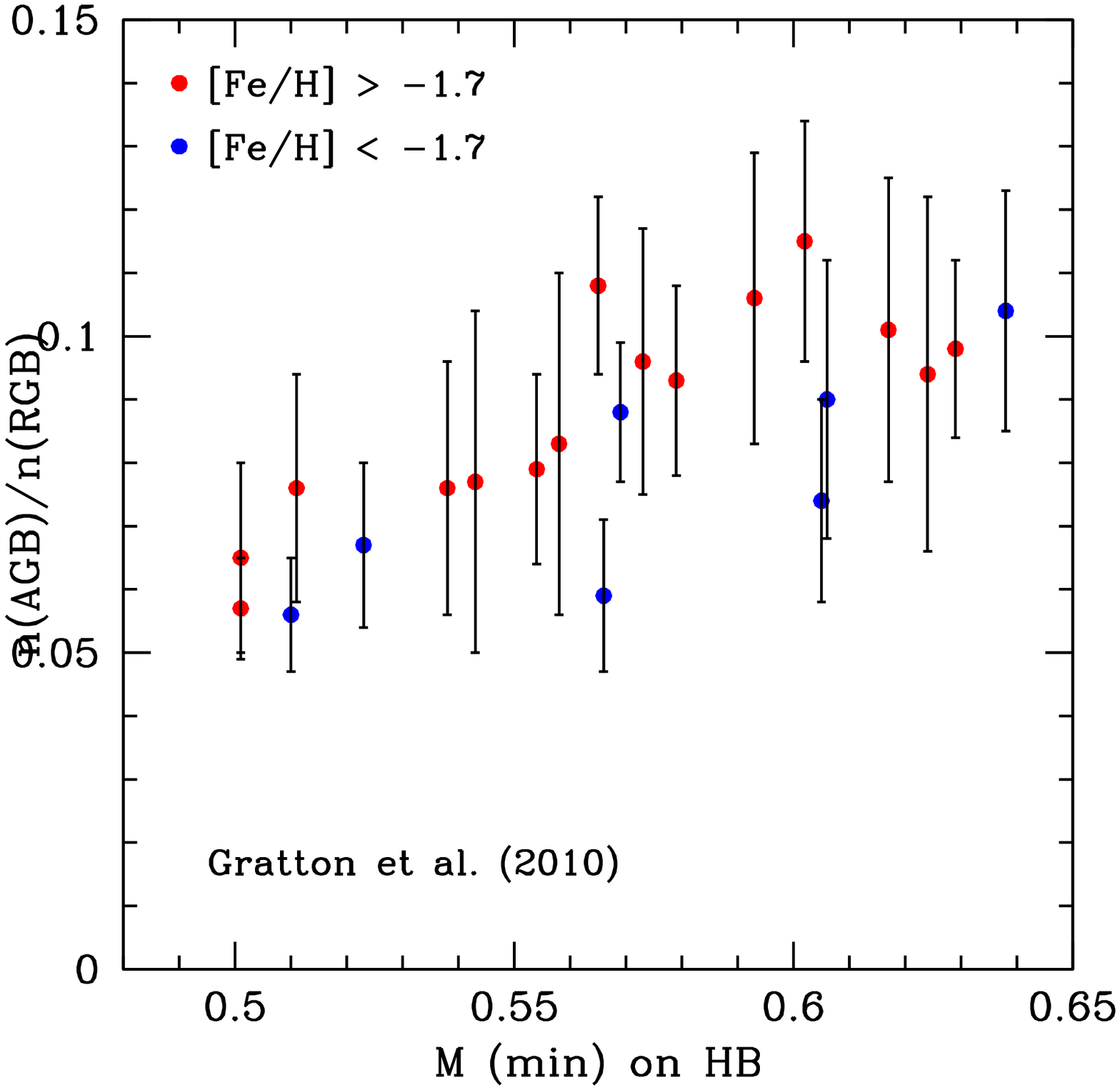}
\caption{As in the previous figure, but taking only Galactic field stars 
from Carretta (2013) within $\pm2 \sigma$ from the average Na at any 
metallicity. AGB stars from several studies are shifted to match the minimum Na
(left panel).In the right panel, the number ratio of AGB to RGB stars is plotted
as a function of the minimum mass on the HB from Gratton et al. (2010a).}
\end{center}
\end{figure}

In Figure 10  I use the sodium excess with respect to the homogenised sample of 
field stars from Carretta (2013) to show that most RGB stars belong to the
second generation (left panel, cluster stars from our FLAMES survey). I also show
AGB stars from some clusters (right panel) and apparently they are more 
first generation.
By taking only field stars within 2$\sigma$ from the average at any metal
abundance (Fig. 11), we can shift all
samples of AGB stars in GCs on the same scale, regardless of details in the
original analyses. It is evident that some clusters show a large number of second
generation AGB stars, others only a few (Figure 11, left panel).
The correlation between the number ratio in AGB and the minimum mass on the HB
(Gratton et al. 2010a, reproduced in Figure 11, right panel) tells us that the
lack is not an {\it in situ} process, but depends on stars failing to reach the AGB
phase (see Greggio and Renzini 1990), likely due, again, to their enhanced He 
abundance.

\bigskip
The issue of multiple stellar populations in the AGB phase is however still 
open. For example, if we take the AGB failure
rate defined by MacLean et al. (2016) from the fraction of SG stars in AGB and 
RGB, the few SG AGB stars in NGC 6752 are explained only with unlikely enhanced mass
loss on the HB (see Campbell et al. 2013, Cassisi et al. 2014).
On the other hand, the rate in NGC 2808 contrasts with the expectation that a
relevant fraction of HB stars should miss entirely AGB due to low mass and
enhanced Helium (e.g. D'Antona et al. 2005), resulting into different
distributions in RGB and AGB, that are not observed (Wang et al. 2016).

Several factors contribute to make the situation unclear for AGB stars:
observations are done using only AGB samples or through a comparison of both 
AGB and RGB; 
different species are used to tag SG stars (Na only, Na and O, Al only, CN only) 
and, as we will see below, different elements are not always coupled (e.g. Smith
et al. 2013); NLTE effects are more relevant in AGB stars; and so on.

On the theoretical side, it is not clear if different scenarios are able to fully
explain the observations. The difference for NGC 2808 and 6752 is explained
because of the dependence of the maximum Na dispersion in AGB on age or 
metallicity in the FRMS scenario (Charbonnel and Chantereau 2016). However, by
interpolating in the age-metallicity plane, for each reasonable combination
age-metal abundance one derives approximatively the same maximum Na dispersion in
AGB. In the massive AGB scenario, on the other hand, to different ranges in Na
must correspond similar values of He, which is provided by the second dredge-up.
Something seems to be still missing, in order to explan the observations in this
phase.

However, as a first summary, we can state some firm points. The chemical
signatures observed in at least two stellar generations in GCs are well matched
by hot H-burning nucleosynthesis that perfectly reproduces all the observed
(anti)correlations, provided that a mandatory amount of pristine gas is
available to dilute nuclearly processed ejecta.
The so formed different stellar generations follow rather well the stellar
evolution requirements, with SG stars nicely segregated according
to their He content, on the HB and probably also on the AGB.

\section{In what stellar systems do we see evidence of multiple populations?}
By using again mainly the Na-O anticorrelation, we may ask what are the
stellar systems where we see the multiple populations phenomenon? 

In almost every halo, bulge and disk Milky Way globular cluster, over
two order of magnitude in total mass, from the tiny Pal 5 in
dissolution (Smith et al. 2002) up to $\omega$ Centauri (Johnson and Pilachowski
2010, Marino et al. 2011c), a likely remnant of a dwarf galaxy (Figure 12).

\begin{figure}
\begin{center}
\includegraphics[scale=0.50]{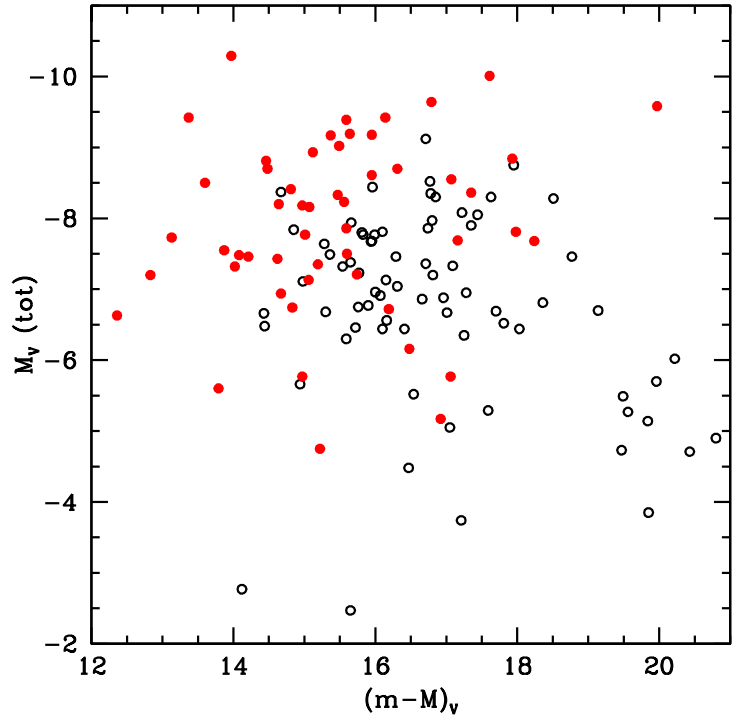}
\caption{the absolute magnitudes of GCs are plotted as a function of the distance
modulus from Harris (1996, 2010 edition). Filled circles are GCs with observed
Na-O anticorrelation. The plot is updated at August 2016.}
\end{center}
\end{figure}

In clusters (suspected to be) of extragalactic origin, like M~54, NGC~1851,
NGC~1904, NGC~2808, M~68 (see Bellazzini et al. 2008, Forbes and Bridges 2010),
all with the Na-O anticorrelation studied in our FLAMES survey, and in true, 
old extragalactic clusters belonging to the Large Magellanic Cloud or to the
Fornax dwarf spheroidal galaxy (Figure 13, from Johnson et al. 2006, Letarte et
al. 2006, Mucciarelli et al. 2009). Note that no evidence of self-enrichment was
found in the $intermediate$ age LMC clusters by Mucciarelli et al. (2008).

\begin{figure}
\begin{center}
\includegraphics[scale=0.40]{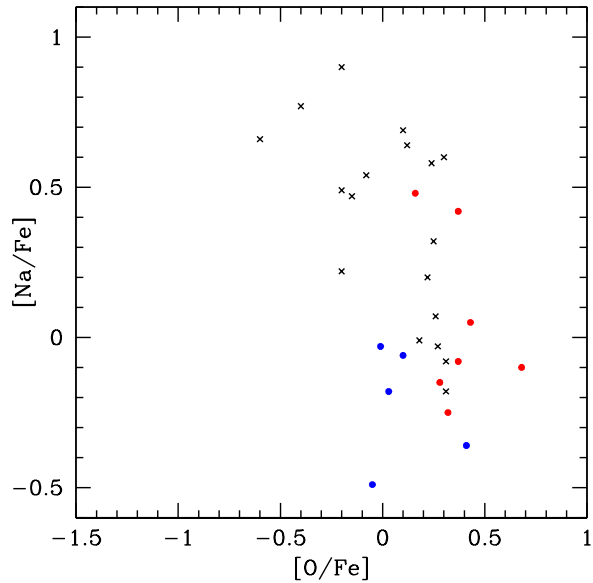}
\caption{Na-O anticorrelation in stars of extragalactic GCs: 4 old GCs in LMC
(Johnson et al. 2006: blue circles), 3 GCs in the Fornax dSph (Letarte et al.
2006, red circles), and 3 old GCs in LMC (Mucciarelli et al. 2009, black
crosses).}
\end{center}
\end{figure}

In so called {\it iron complex} clusters, where the Na-O anticorrelation is 
found in $each$ metallicity component (Figure 14): from those associated to nuclei of 
present or former dwarf galaxies, M 54 (Carretta et al. 2010a,e) and 
$\omega$ Centauri (e.g. Norris and Da Costa 1995, Johnson and Pilachowski 2010,
Marino et al. 2011c) to the normal clusters with small or moderate dispersion in
iron and heavy elements like NGC~1851 (Carretta et al. 2011) or M~22 (Marino et
al. 2011a). This is a currently increasing sample (see M~2, Yong et al. 2014;
NGC~5286, Marino et al. 2015; NGC~5824, Roederer et al. 2016; M~19, Johnson et
al. 2015b).

Dwarf galaxies show by definition a metallicity dispersion, but the
anticorrelation is only found in their globular clusters, not in the field (e.g.
Carretta et al. 2010e). 
Of course the progenitors of M 54 and $\omega$ Centauri were much more close to the
central regions of the Milky Way than present-day dSphs.

The Na-O anticorrelation is found in clusters both  likely accreted or formed 
in situ in the Milky Way, found to lie
along two different age-metallicity relations (see VandenBerg et al. 2013,
Leaman et al. 2013).
In other words this anticorrelation, the main chemical signature of multiple stellar
populations, is widespread among clusters and very likely related to
their origin since these self-enrichment events occurred in the first 1\% of
their lifetime and was proposed (Carretta et al. 2010b) to be considered as 
the main signature of a genuine globular cluster.

\bigskip
One of the best example/application can be borrowed from Doug Geisler. If one
looks at an image of Pal 5 it is easy to think of it as an open cluster or an 
association, whereas the image of NGC~6791 shows a beautiful globular form.
However, according to the presence of abundance variations, Pal 5 has multiple
populations and it is a proper globular cluster, NGC~6791 is not (Bragaglia et al.
2014, Cunha et al. 2015), and it is only a 
strange old, metal-rich open cluster (but see Geisler et al. 2012 for a
different view on this object).
As a safety check, open clusters follow the pattern of field stars and show no
sodium excesses anticorrelated to oxygen depletions (Figure 15, see De Silva et al.
2009, MacLean et al. 2015, Bragaglia et al. 2014).

\begin{figure}
\begin{center}
\includegraphics[scale=0.50]{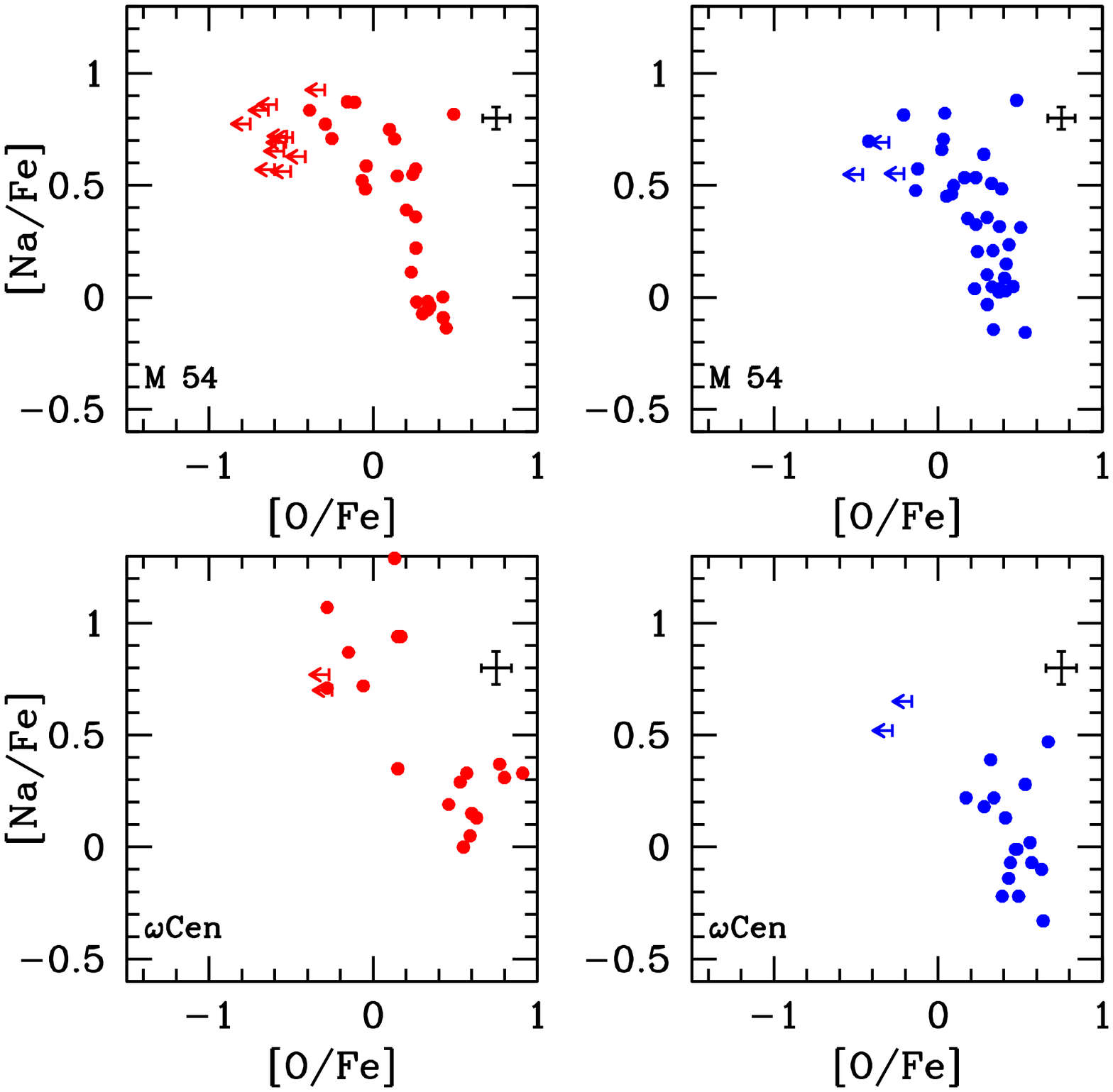}
\includegraphics[bb=19 388 539 680, clip, scale=0.50]{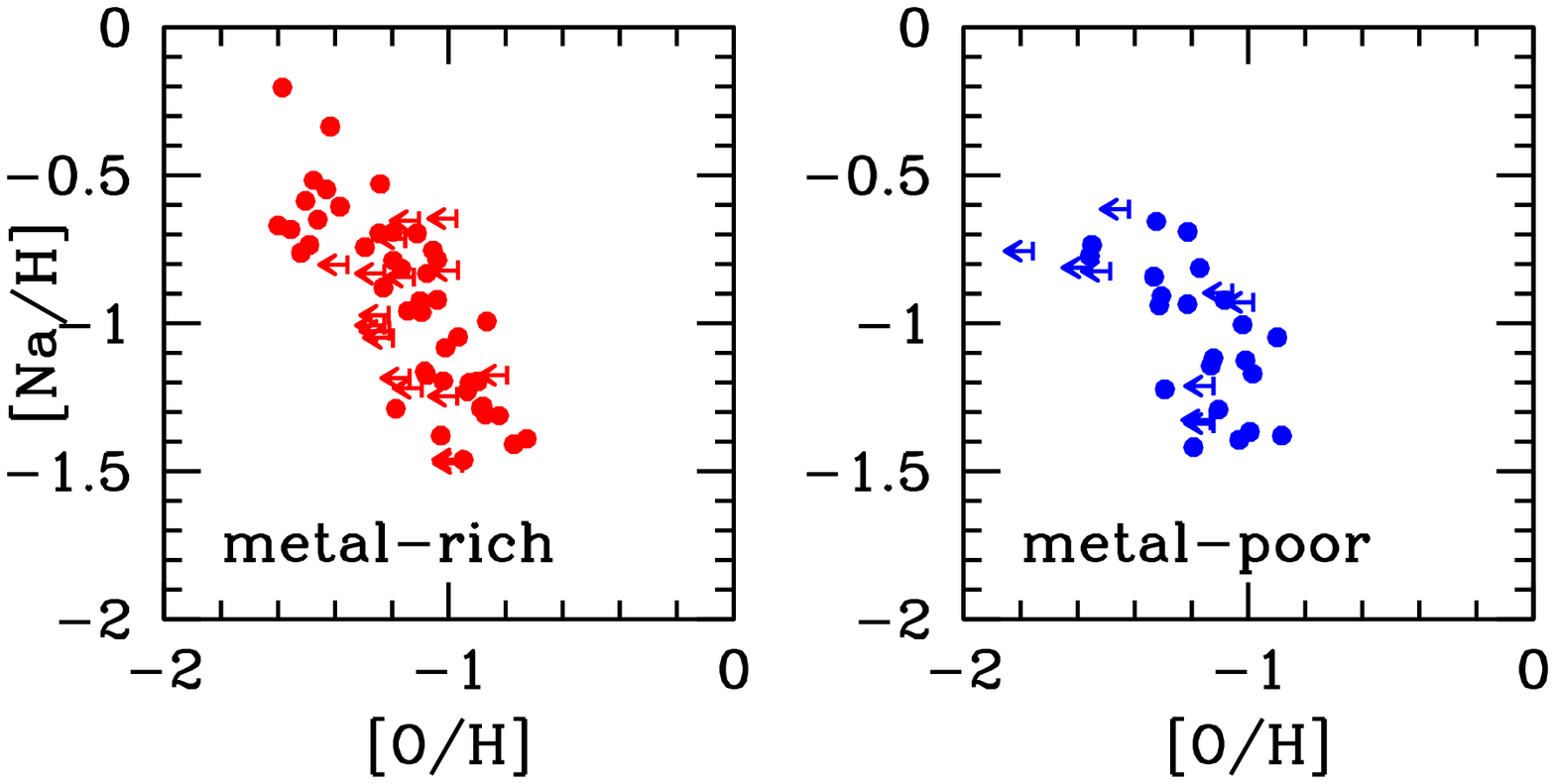}
\includegraphics[bb=19 388 539 680, clip, scale=0.50]{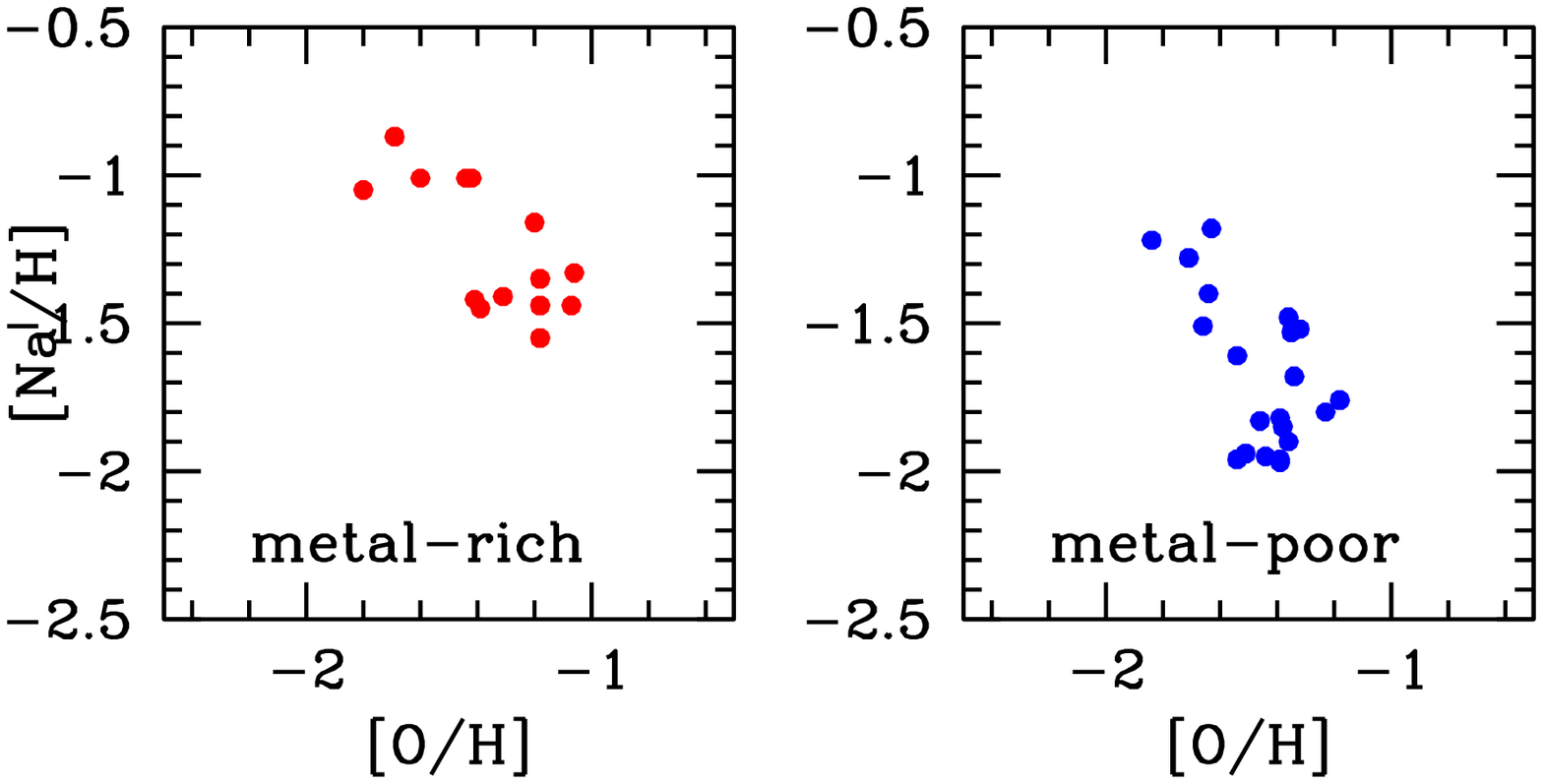}
\caption{Na-O anticorrelation in GCs with metallicity dispersion. First row:
M~54 (Carretta et al. (2010a,e); second row: $\omega$ Cen (Norris and Da Costa
1995); third row NGC~1851 (Carretta et al. 2011), fourth row: M~22 (Marino et
al. 2011).}
\end{center}
\end{figure}

\begin{figure}
\begin{center}
\includegraphics[scale=0.50]{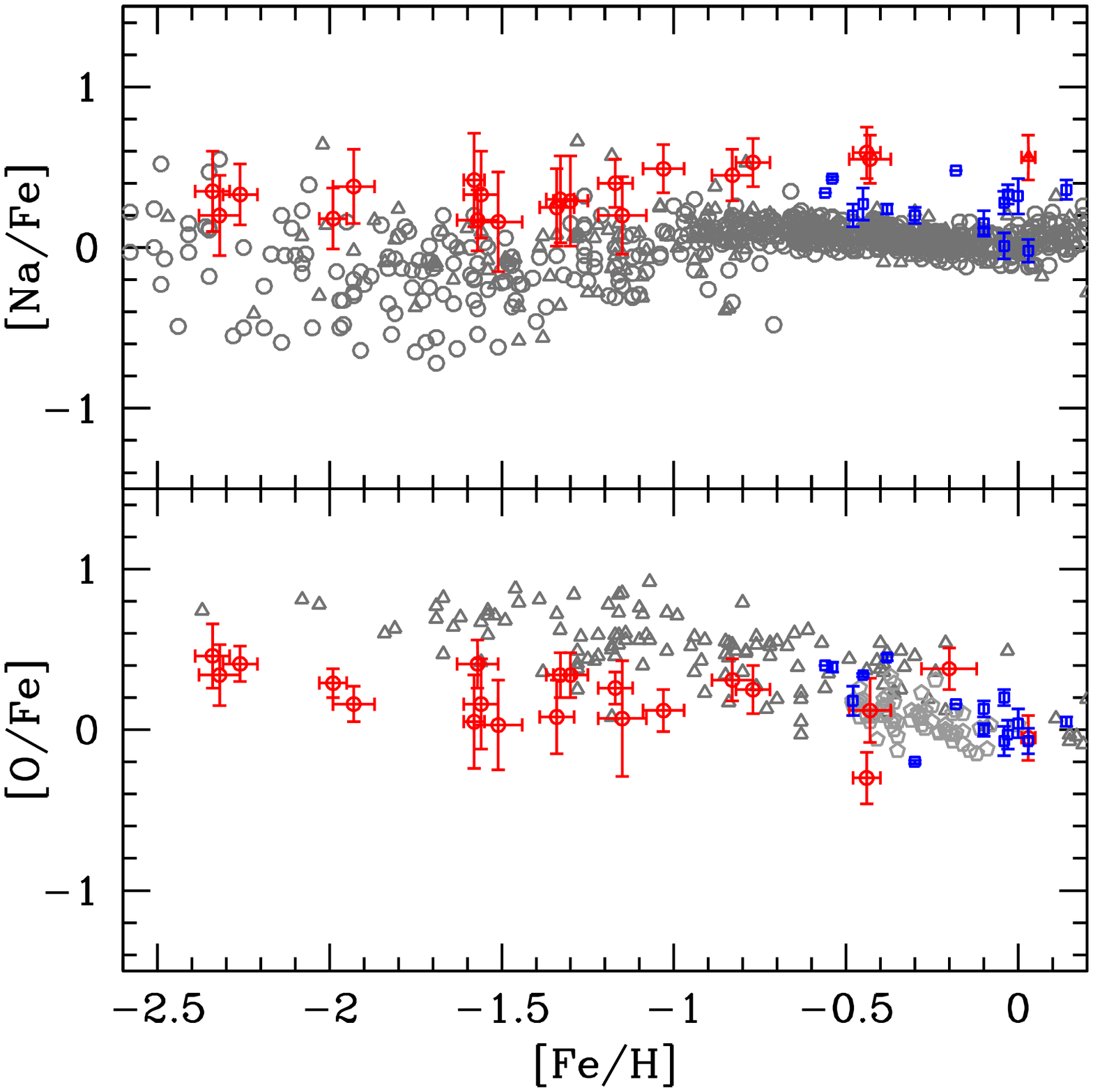}
\caption{[Na/Fe] and [O/Fe] ratios in Galactic field stars as a function of
metallicity. Red circles indicate the average O and Na content in GCs from
Carretta et al. (2009b) and blue squares are the same for open cluster from De
Silva et al. (2009).}
\end{center}
\end{figure}

\section{What is the relation with global cluster properties?}
Finally, what is the relation between the multiple stellar populations in GCs 
and the global cluster properties?

With good statistics we can quantify the extent of multiple populations using
the interquartile range of the oxygen-sodium ratio, IQR[O/Na], as suggested by
Carretta (2006).
We found that the total mass is the driving parameter: in Figure 16 (upper
panel). we see that a very significant correlation exists between the extension
of the anticorrelation and the total  mass of the clusters. We used as a proxy
for the mass the total cluster absolute magnitude, so that this is the
present-day mass, including the initial mass and the memory of all the internal
and external  processes of dynamical evolution.

\begin{figure}
\begin{center}
\includegraphics[scale=0.50]{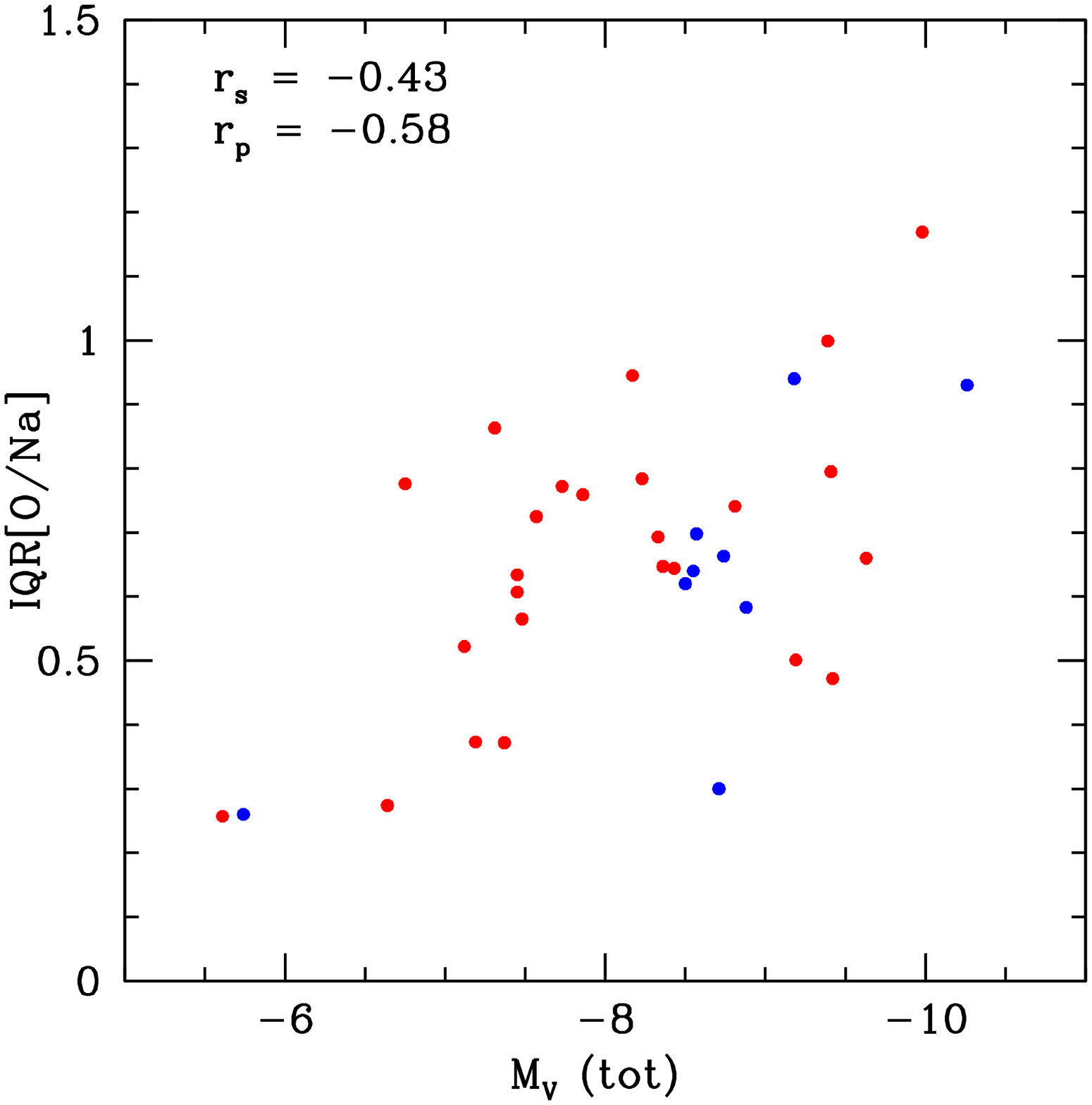}
\includegraphics[scale=0.50]{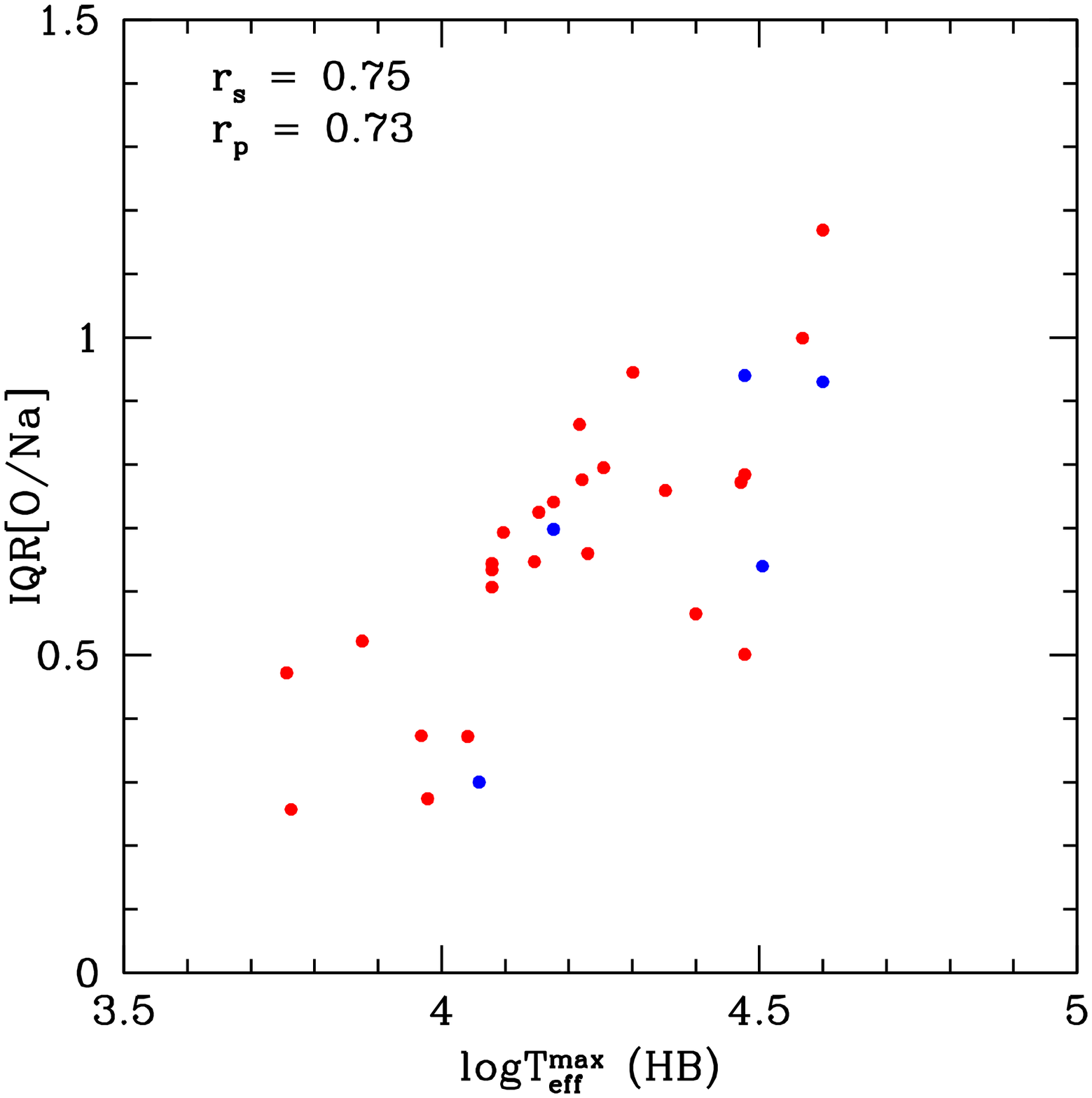}
\caption{interquartile range of the [O/Na] ratio as a function of
the total absolute magnitude of GCs (upper panel) and of the maximum temperature
along the HB from Recio-Blanco et al. (2006, lower panel). Spearman and Pearson
coefficients for the linear regression are indicated.}
\end{center}
\end{figure}

Moreover, the link with the main product of H-burning, helium, predicts that 
helium enhanced stars populate bluer and hotter locations on the HB (e.g.
D'Antona et al. 2002, Catelan 2009, Gratton et al. 2010b), since
He-enhanced stars have slightly smaller evolving masses, and in fact
we found that multiple populations are related to the HB morphology.
As predicted, clusters with more extreme anticorrelations have
bluer and hotter HBs (Carretta et al. 2007d, Carretta 2015 and references
therein), as shown in Figure 16 (lower panel).

\bigskip
Since the minimum sodium in GCs nicely follows the pattern of field stars, 
from spectroscopy we can select with accuracy the primordial component in each cluster and so 
also the remaining fraction of second generation stars. In figure 17 (upper
panel), the P component includes all stars along the Na-O anticorrelation
with Na abundances [Na/Fe]$<$ [Na/Fe]$_{min}$ +4$\sigma$.

In present day clusters second generation stars are the majority (Figure 17,
lower panel), apart from
a few cases like M~53 (Boberg et al. 2016) or NGC~6723 (Gratton et al. 2015).
However, SG stars are formed by only a fraction (the most massive stars) of the
FG, and this is known as the so called {\it mass budget problem}, usually 
solved by assuming that the precursors of the GCs were much more massive than 
present-day GCs (Bekki et al. 2007, and many others), losing about 90\% of their FG stars and becoming good
candidates for main contributors to the halo (see e.g. Carretta 2016).

\begin{figure}
\begin{center}
\includegraphics[scale=0.50]{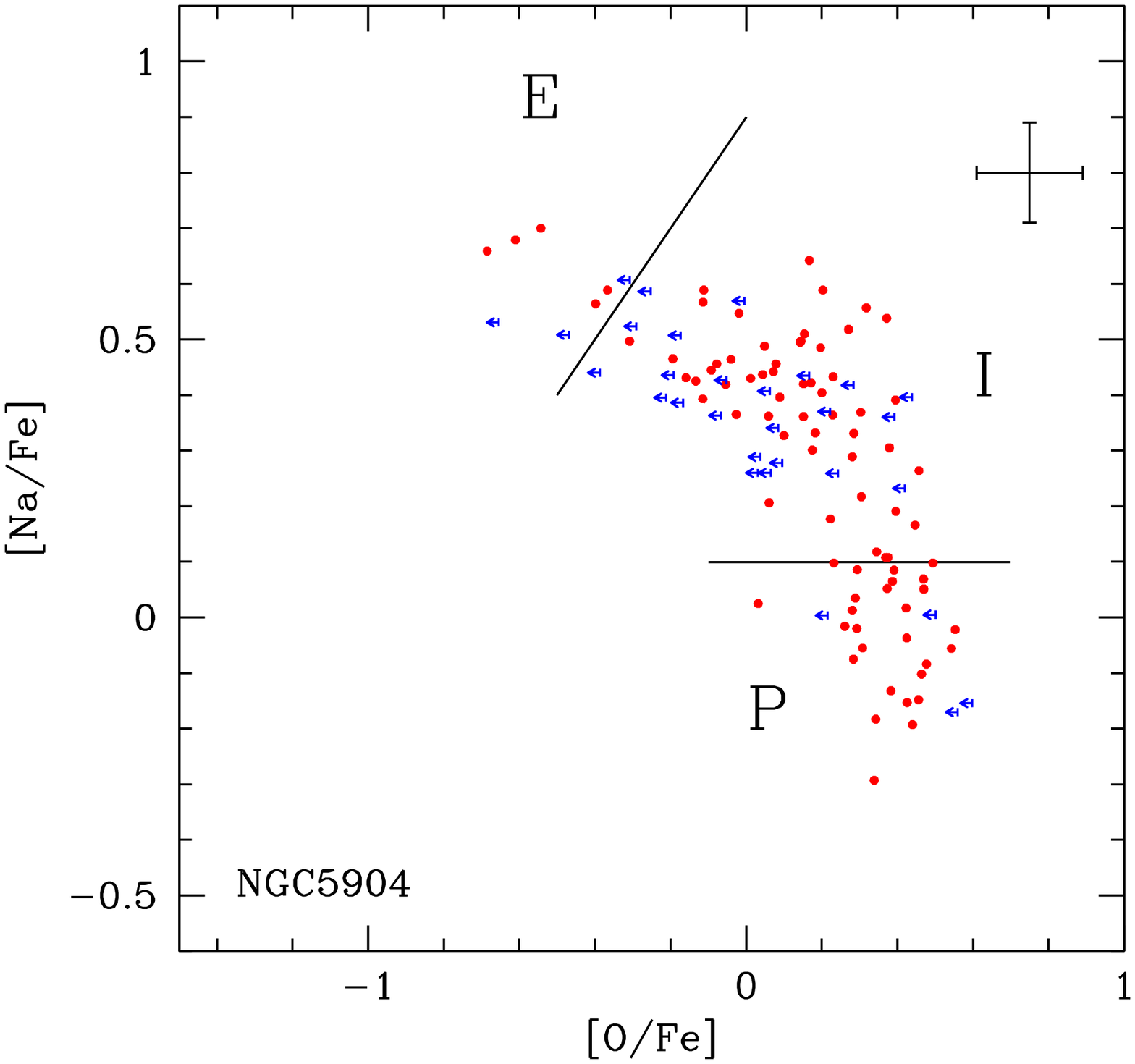}
\includegraphics[scale=0.50]{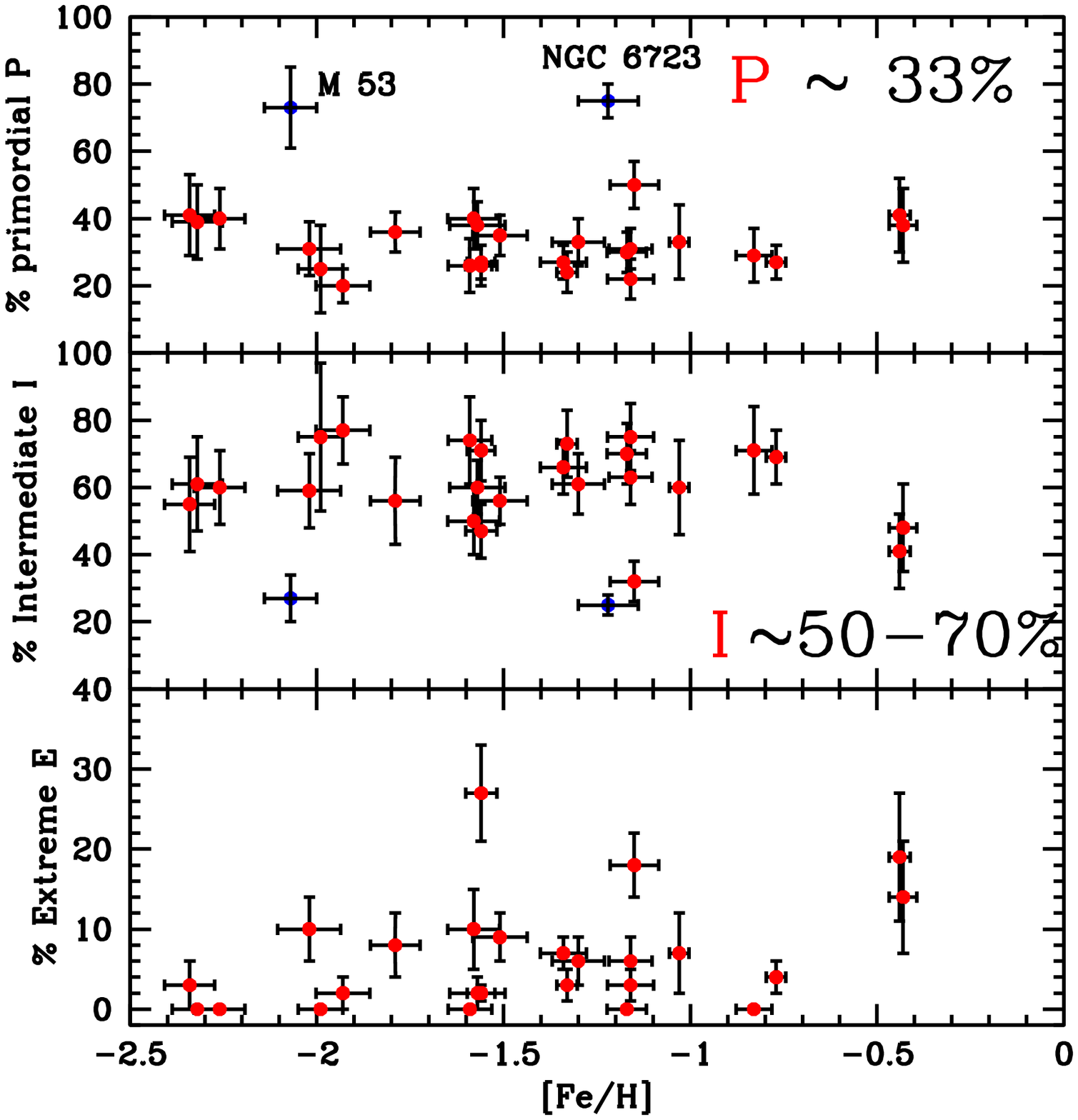}
\caption{Division of RGB stars in NGC~5904 into the primordial P component of FG
and into the intermediate I and extreme E component of SG according to their O,
Na abundances (upper panel). In the lower panel  the fractions of
P,I,E stars in GCs of our FLAMES survey (Carretta et al. 2009a; red points) and 
for M~53 (Boberg et al. 2016) and NGC~6723 (Gratton et al. 2015), in blue, are plotted.}
\end{center}
\end{figure}

\begin{figure}
\begin{center}
\includegraphics[scale=0.40]{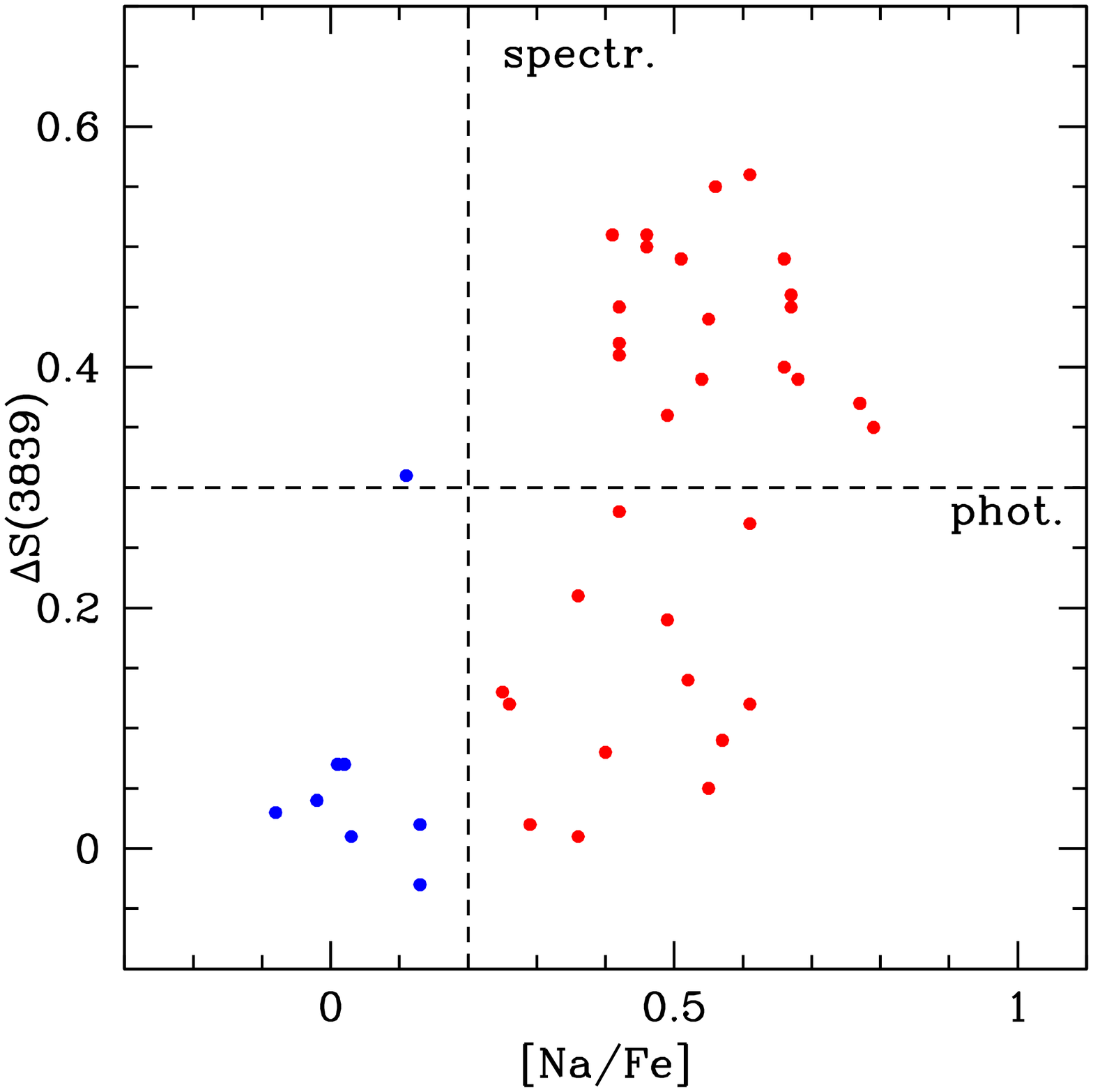}
\includegraphics[scale=0.40]{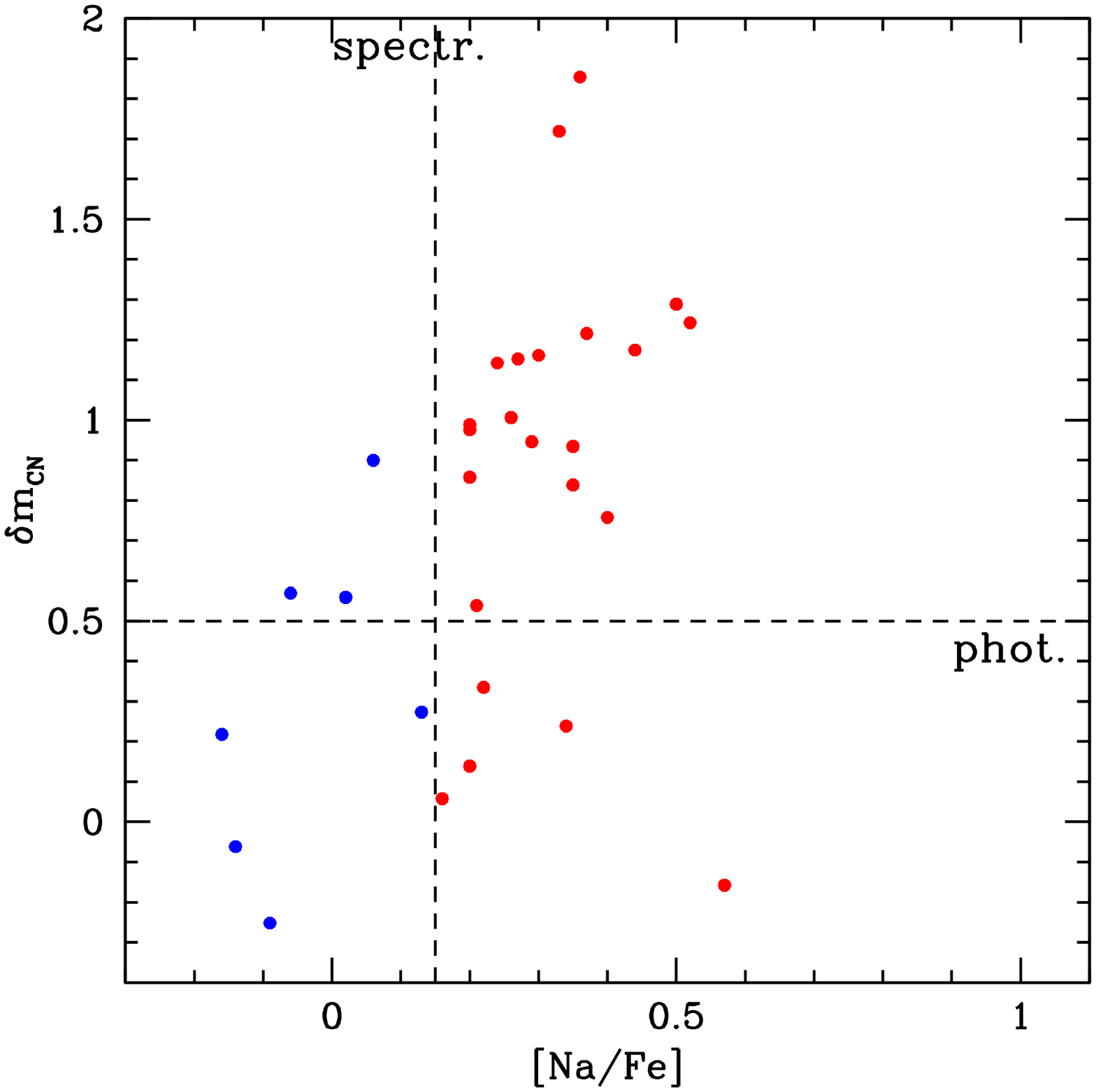}
\includegraphics[scale=0.40]{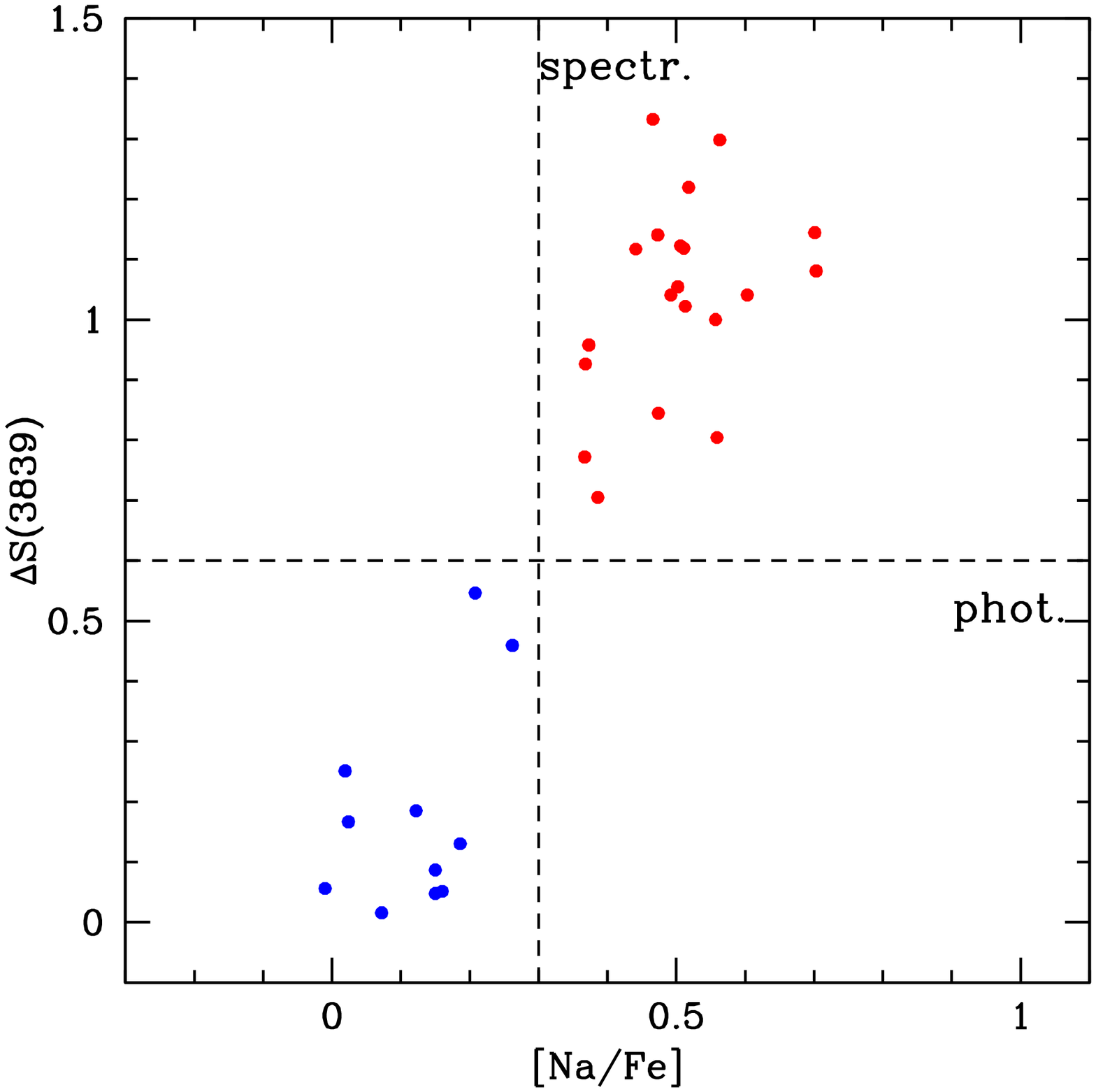}
\caption{Upper left panel: $\lambda 3883$ CN residual $\Delta S(3839)$ with respect to the
baseline of CN-weak stars (Smith et al. 2013) as a function of [Na/Fe] ratios 
(Carretta et al. 2009a) in M~5. The horizontal line is the population division
based on CN/photometry, whereas the vertical one indicates the ratios FG/SG
following the spectroscopic criterion by Carretta et al. (2009a). Upper right
panel: the same for M~13 from Smith and Briley (2006) for CN and Johnson and
Pilachowski (2012) for Na. In the lower panel, the same for M~4 from Smith and
Briley (2005) for CN and Marino et al. (2008) for Na.}
\end{center}
\end{figure}

\begin{figure}
\begin{center}
\includegraphics[scale=0.40]{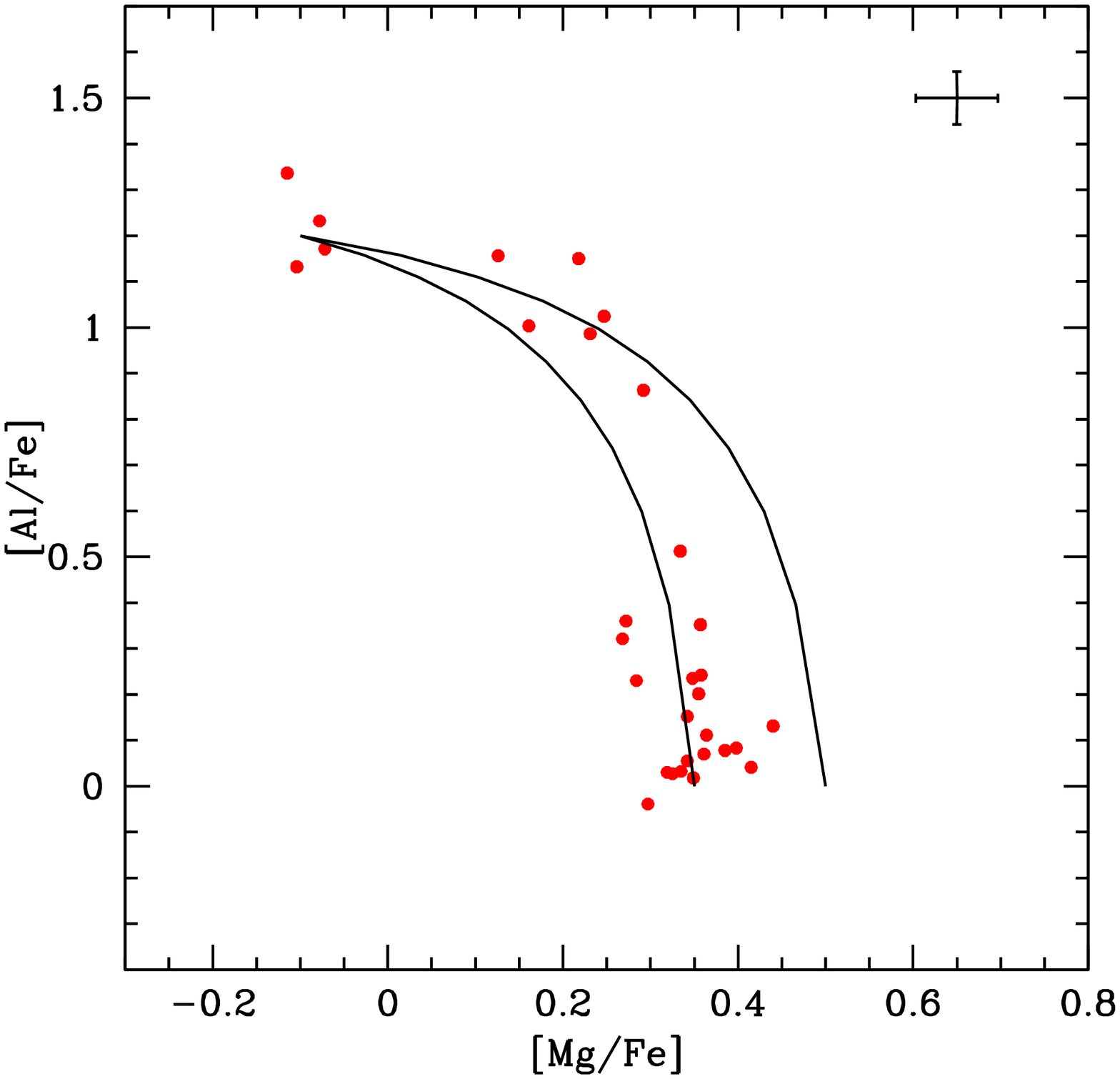}
\includegraphics[scale=0.40]{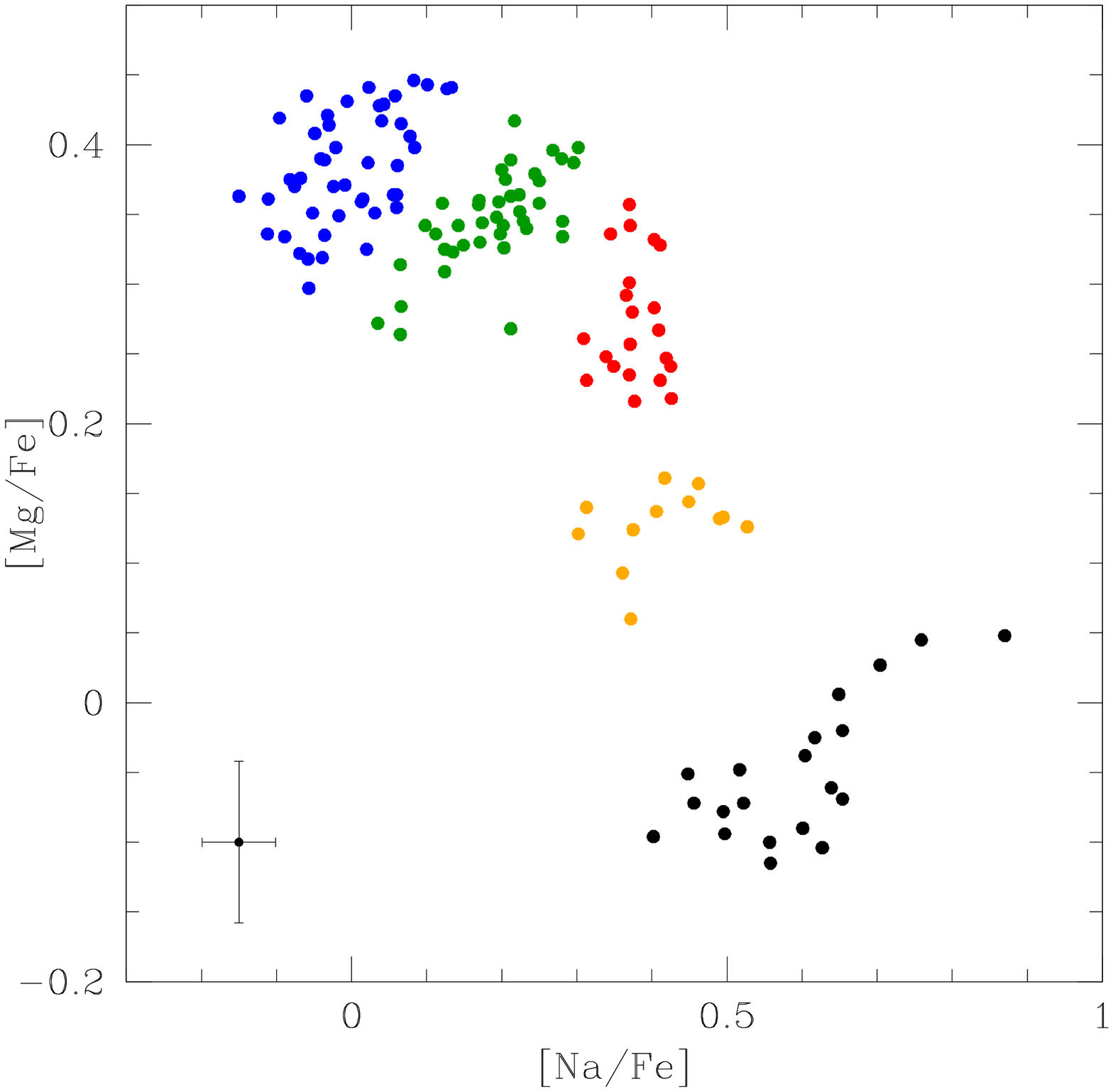}
\caption{[Al/Fe] ratios as a function of [Mg/Fe] ratios in NGC~2808 (Carretta
2014, left panel) with superimposed two dilution models. No single dilution
model is able to fit all the three components. Right panel: five groups with
distinct chemical composition on the RGB of NGC~2808 from Carretta (2015).}
\end{center}
\end{figure}

A selective mass loss is both predicted by different scenarios and dynamical
simulations (e.g. D'Ercole et al. 2008, Decressin et al. 2007, Vesperini et al.
2013) and supported by the observation that second generation stars are usually
more concentrated (see e.g. Lardo et al. 2011, Milone et al. 2012, Carretta
2015), although there are some counterexamples (e.g. Larsen et al. 2015).

\bigskip
These basic assumptions must confront with two major challenges. 

One is the so-called {\it external mass budget problem} because
evidence is accumulating that in dwarf galaxies GCs cannot have
been more massive than 4-5 times, initially, as they account for about 25\% of
the galaxy mass in metal poor stars (Larsen et al. 2012, 2014b, Tudorica et al.
2015).

Furthermore, simulations of mass loss from gas expulsion predict a strong
anticorrelation between the fraction of SG stars and the cluster mass, that is
not observed (Khalaj and Baumgardt 2015), and no variation of the fraction of SG
as a function of cluster mass or  galactocentric distance or metallicity is
observed (Kruijssen 2015, Bastian and Lardo 2015), at odds with what expected 
by mechanisms of mass loss.

\bigskip
Just a caveat: it is risky to mix estimates of the population ratios from 
spectroscopy, photometry and simulations, as in the last study, because of the 
decoupling between CN and heavier elements already noted by Graeme Smith and
collaborators (see Smith et al. 2013, Smith 2015). 

Photometry on one hand and spectroscopy on the
other hand are not seeing exactly the same things, because most photometric
indexes are essentially sensitive only to CN, NH or CH features. As a consequence of
the temperature involved in the H-burning reactions, they sample different 
range of temperatures, and therefore of polluter masses, than spectroscopy with
the observations of the outcome of the NeNa and MgAl cycles.

Usually a division based on CN derives a 50-50 population ratio (e.g. Figure
18, upper left panel for M~5). However, adopting our criterium (Carretta et al.
2009a) based on Na you find that some SG stars are instead counted as FG by CN properties
or photometry,
even if in these stars Na is as high as in SG stars. Moreover, this discrepancy
seems to vary from cluster to cluster (see the cases for M~13 and M~4, upper
right and lower panels, respectively, in Figure 18).

Finally, it appears to be difficult to account for all the combined spectroscopic
and photometric evidence within a single scenario (e.g. Bastian et al. 2015).
In particular, the discrete components seen by spectroscopic observations 
in some clusters (NGC~6752, Carretta et al. 2012a; NGC~2808, Carretta 2014
2015, see Figure 19) suggest that a dilution model with only one class of
polluters could not be enough for all components.

So, in conclusion taking into account the large variety of chemical signatures
in GCs (spread in light elements, always; also in heavier elements and
metallicity, in some cases) the most honest summary I can give you from the 
spectroscopic evidence is this: a genuine, {\it bona fide} old globular cluster
is a system where a chain of events (not yet well understood) concurred to form
at least two stellar generations with small age differences and huge variations
in the content of products of hot H-burning nucleosynthesis.

\section{Acknowledgements}
I would like to thank the organizers for inviting me. This was a mixed audience,
with different expertises, so that we could learn from each others. I wish to
thank in particular Michela Mapelli and Elena Sabbi for their nice review and
summary talks, respectively.

\end{document}